\documentclass[prd,tightenlines,nofootinbib,amsfonts,amssymb,amsmath,color,11pt]{revtex4}
\usepackage{graphicx}	
\usepackage{color}

\usepackage{amsfonts}
\usepackage{amsmath}
\usepackage{amssymb}
\usepackage{hyperref}

\newcommand{\dif}{\mathrm{d}}

\def\be{\begin{equation}} 
\def\ee{\end{equation}}
\def\bea{\begin{eqnarray}}
\def\eea{\end{eqnarray}}
\def\nn{\nonumber}

\begin{document}

\title{Subleading Effects  and the Field Range in Axion Inflation }
\author{Susha Parameswaran$^a$, Gianmassimo Tasinato$^b$, Ivonne Zavala$^b$}

\affiliation{$^a$ Department of Mathematical Sciences, University of Liverpool,
Liverpool, L69 7ZL, UK \\
$^b$  Department of Physics, Swansea University, Swansea, SA2 8PP, UK }

\begin{abstract}

 An attractive candidate for the inflaton is an axion slowly rolling down a flat potential protected by a perturbative shift symmetry.  
 Realisations of this idea within large field, natural and monomial inflation have been disfavoured by observations and are difficult to embed in string theory.  We show that subleading, but 
  significant non-perturbative  corrections can superimpose sharp cliffs and gentle plateaus into the potential, whose overall effect is to enhance the number of e-folds of  inflation.
   Sufficient e-folds are therefore achieved for smaller field ranges compared to the potential without such corrections.  Thus, both single-field natural and monomial inflation in UV complete theories like string theory, can be restored into the favour of current observations, with distinctive signatures.  Tensor modes result  un-observably small, but there is a large negative running of the spectral index.  Remarkably, natural inflation can be achieved with a single field whose axion decay constant is sub-Planckian.

\end{abstract}

 \maketitle

 \section{Introduction}

The latest results from Planck and BICEP2/Keck \cite{Planck15Infla,BICEP15} are in  agreement with the simplest inflationary scenario, driven by a single scalar field slowly rolling down a very flat potential.  The conditions on the inflaton potential for a sufficiently long epoch of slow-roll inflation are:
\be
\epsilon_V \equiv \frac{M_{Pl}^2}{2}\left( \frac{V'}{V}\right)^2 \ll 1 , \quad |\eta_V| \equiv M_{Pl}^2  \frac{|V''|}{V} \ll 1 \,. \label{E:potSR}
\ee
The challenge in slow-roll inflation is that higher order corrections to the effective potential generically steepen the potential and spoil the slow-roll conditions.  Higher dimensional operators, $\mathcal{O}_d = c_d V(\phi) (\phi/M_{Pl})^{d-4}$, typically lead to order one contributions to $\eta_V$ at dimension six.  Moreover, for even moderately large $\phi \gtrsim M_{Pl}$, Planck suppressed operators are dangerously large, and can even prevent inflation from occurring at all.   

An attractive way to address this challenge is to invoke a symmetry that forbids large quantum corrections. One possibility  is the shift symmetry enjoyed by the currently favoured Starobinsky or Higgs inflation, in the limit of large values for the inflaton.  Alternatively, the shift symmetry governing axions provides such a symmetry.  The original realisation of this idea is  natural inflation \cite{NI}.  Here, the continuous shift symmetry of the axion is broken non-perturbatively to a discrete symmetry, leading to a potential of the form:
\be
V(\phi) = \Lambda^4 \left( 1 - \cos\left(\frac{\phi}{f} \right) \right) \,,
\ee
with $f$ the axion decay constant.  The potential is sufficiently flat for slow-roll, (\ref{E:potSR}), for $f \gtrsim 4 M_{Pl}$, which also suggests super-Planckian field ranges.  A second class of natural models is axionic realisations of monomial inflation \cite{chaotic}, whose attempted string theoretic embeddings are known as axion monodromy \cite{AM1,AM2,KS,FtermAM}.  
  Here the axion  shift symmetry is broken e.g. spontaneously via its coupling to some non-trivial background flux, leading to a simple power law potential, for example the  quadratic potential:
\be
V(\phi) = \frac12 m^2 \phi^2 \,.
\ee
The axion decay constant can now be sub-Planckian.  But again, slow-roll conditions, (\ref{E:potSR}), are satisfied only for super-Planckian field ranges, $\Delta \phi \approx 15 \, M_{Pl}$.  
 See \cite{Pajer:2013fsa} for a review. 

Large field models present both observational and theoretical challenges.    Assuming slow-roll from the time the observed perturbations in the CMB exited the horizon up to the end of inflation, field ranges are related to the amplitude of tensor modes via the Lyth bound \cite{Lyth,Lotfi,GRSZ2}:
\be
\frac{\Delta \phi}{M_{Pl}} \gtrsim 2 \times \left(\frac{r}{0.01}\right)^{1/2} \,.
\ee 
Indeed, both vanilla natural and $\phi^2$ models predict a tensor to scalar ratio, $r \sim 0.1$, and are thus in tension with the most recent bounds from BICEP2/Keck \cite{BICEP15}, $r < 0.07$.  Furthermore, a large amplitude of primordial gravitational waves implies that the inflationary energy scale is high, via the relation
\be
V_{inf}^{1/4} \simeq  \left(\frac{r}{0.1}\right)^{1/4}\times 1.8 \times 10^{16} \textrm{ GeV} \,.
\ee
UV completions of high scale inflation via string theory are difficult, due to the proximity between $V_{inf} \sim 10^{16}$ GeV and the string scale, which is typically $M_s \lesssim 10^{17}$ GeV for perturbative string theory \cite{KPZ, BMc}.  This proximity puts under pressure the validity of 4D effective field theory during inflation, where, in order to be able to neglect massive string excitations and Kaluza-Klein modes, a hierarchy $V_{inf}^{1/4} \lesssim M_{kk} \lesssim M_s$ is required,  $M_{kk}$ being the compactification scale.  Notice that this tension between high inflationary scale and perturbative string theory also renders it hard to consistently embed the   observationally favoured Starobinsky or Higgs inflation into string theory.  Finally, there have also been suggestions that large axion decay constants are not possible in theories of quantum gravity like string theory \cite{WGC}\footnote{However,
 counterexamples to this claim do exist,  both in single field \cite{AZ, KT, KPZ} and multifield models \cite{KNP,Nflation}.}.

At the same time, it has long been known  that axion models give vanilla natural or monomial inflation only to leading order.  Whilst the effective potential is protected from dangerous perturbative corrections, non-perturbative effects like instantons do give further contributions of the form $\sum_n \Lambda_n^4 \cos(n\phi/f)$, where $\Lambda_n$ are  mass scales.  On the one hand, these higher harmonics have been argued to generically dominate for large axion decay constants.  Then large and frequent modulations are introduced into the potential, trapping the inflaton in a local minimum before it reaches the global one, and obstructing large numbers of e-folds of slow-roll natural inflation \cite{Banks}.   On the other hand, when non-perturbative corrections correspond to tiny, frequent superimposed features in the slow-roll potential, their impact on the background trajectory of the inflaton is negligible, whilst leaving only small imprints on the CMB.  Indeed, modulations have been shown to lead to resonant enhancement of the bispectrum \cite{Chen:2008wn, nonGAM,FP}, a large, possibly oscillating running of the scalar spectral index  \cite{Kobayashi:2010pz}  
 and a slightly reduced tensor to scalar ratio \cite{Tatsuo, Kappl:2015esy, Choi:2015aem}.

\smallskip

\begin{figure}[!htb!]
\begin{center}
\includegraphics[width=0.47\textwidth]{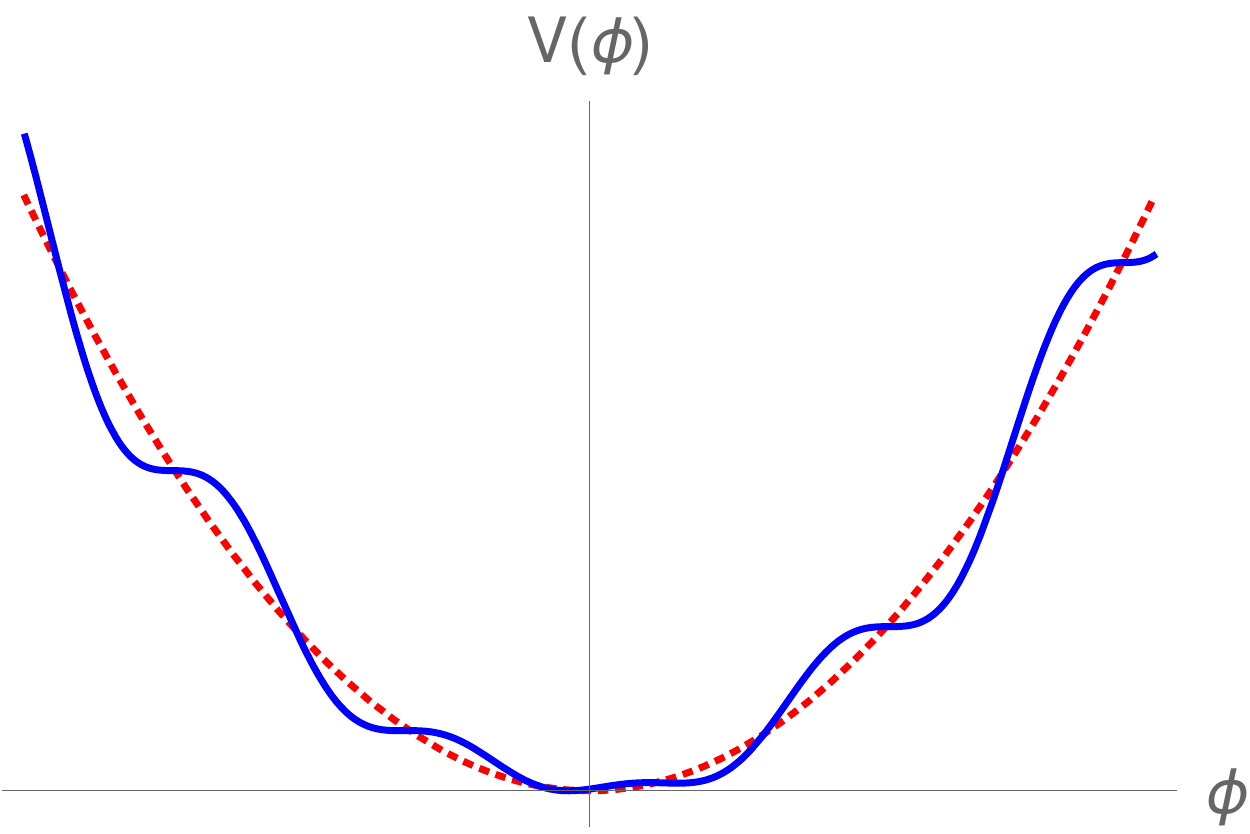}
 \caption{A typical potential for the axion before (red, dotted) and after (blue, solid) including subleading, but significant, non-perturbative contributions.  Depending on the parameters, the higher harmonics can introduce smooth step-like structures into the potential, with steep cliffs connected by flat plateaus.}\label{F:Potential}
\end{center}
\end{figure}

As we  show in this paper, non-perturbative corrections to axion models of inflation  not only allow  inflation, but can even help it.  In particular, when non-perturbative corrections are subleading, but significant, the inflaton potential can acquire bumps which take the form of sharp cliffs and gentle plateaus (see Fig.~\ref{F:Potential}).  These features can modify the inflaton background 
  trajectory, in such a way as to increase the total number of e-folds (although the dynamics does not generically 
 satisfy slow-roll conditions through the entire inflationary period).  As we will see in the analysis of our explicit scenarios, when the inflaton reaches a gentle plateau represented in   Fig.~\ref{F:Potential}, Hubble
 friction is very effective in rapidly reducing the velocity that the inflaton acquires rolling over the rapid steeps. Hence, thanks to the effect of Hubble friction, most of the inflaton trajectory is spent slowly rolling over the gentle
 plateau. This increases the number of e-folds in comparison with the smooth version of the potential, for given initial conditions and field ranges.  
  The slow-roll parameters are small while on the plateaus, but have large variations when the inflaton rolls down the steep slopes.

With the appropriate tunings -- inevitable in inflation -- we find that single field models of $\phi^2$ and natural inflation can achieve sufficient e-folds for much reduced field ranges, $\Delta \phi \sim (1-4) M_{Pl}$.  Moreover, natural inflation can be obtained with (sub-)Planckian axion decay constants, $f \lesssim M_{Pl}$.  This is to be compared with previous models using small modulations in the potential \cite{AM1, Kobayashi:2010pz, Tatsuo, Kappl:2015esy,Choi:2015aem}, which barely modify the slow-roll behaviour and continue to represent large field models, with the tensor to scalar ratio $r \sim 10^{-2}$ and high scale inflation (difficult to embed consistently in string theory).  
Indeed, when bumps are significant as in Fig. \ref{F:Potential}, $r$ can be brought  below current and future bounds,  and back into the favoured region of current observations (and perturbative string theory). 
The other striking signature of the bumpy models is in a large, negative running of the scalar spectral index at all scales probed by the CMB 
 (small, frequent modulations can instead lead to a running that oscillates between positive and negative \cite{Kobayashi:2010pz, Kappl:2015esy, Choi:2015aem}; see also 
 \cite{Turner,Adshead:2010mc,Easther:2006tv,Encieh} for further observational and theoretical perspectives on this observable), with suppressed running of the running, and they motivate improvements in the measurement of these observables and embedding into consistent perturbative string models of axion inflation.  
 
We end the  introduction by quoting the relevant measurements and bounds on the CMB observables from Planck 2015 \cite{Planck15Infla}.  For the scalar perturbations, the Planck analysis gives\footnote{Including the running of the running (but no tensor modes), the Planck analysis gives $n_s = 0.9569 \pm 0.0077$, $\alpha_s=0.001^{+0.014}_{-0.013}$, $\beta_s=0.029^{+0.015}_{-0.016}$.  However, the models we consider have $\beta_s$ suppressed two orders of magnitude below $\alpha_s$.} (at the pivot scale $k_* = 0.05{\rm Mpc}^{-1}$ and 68\%CL for Planck TT+lowP):  
\bea
&&n_s = 0.9655 \pm 0.0062\,, \nn \\
&&\alpha_s \equiv \dif n_s/ \dif \ln k = -0.0084 \pm 0.0082 \,. \label{E:Planck1}
\eea
Including the tensor perturbations, the results are:
\bea
&&n_s = 0.9667 \pm 0.0066 \,, \nn \\
&&\alpha_s \equiv \dif n_s/ \dif \ln k = -0.0126^{+0.0098}_{-0.0087}\,, \nn \\
&&r < 0.168 \quad (r < 0.07  \,\,\, \text{at 95\% CL according to \cite{BICEP15}})\,.\label{E:Planck2}
\eea
\smallskip

The paper is organised as follows. In the next section we consider simple, single field, bumpy monomial and natural inflationary scenarios, motivated by string theory. Analysing the models numerically, we show that the potentials allow for sufficient inflation with much reduced field ranges, although the slow-roll approximation is not always satisfied through the inflationary evolution. Arranging for slow-roll during horizon crossing, we compute the observables for two benchmark models and find them to be consistent with current measurements and bounds, with distinctive signatures.   Section \ref{S:Conclusions} is devoted to a discussion of our results. An appendix, moreover, presents a simple, analytical model of inflation that shares features
with the scenarios of the main text, and that allows one to physically appreciate in a simple set-up   the arguments we develop in the paper.

\section{Bumpy Monomial and Natural Inflation}

In this section we present two benchmark models, which illustrate the effects of the higher harmonics due to non-perturbative axion effects  as discussed in the introduction, for both monomial and natural inflationary scenarios. In particular, we  consider models in which the frequencies and amplitudes of the oscillations are such as to introduce smooth  step-like features into the potential, with steep cliffs and gentle plateaus.  The main consequences
 are a modification of the background inflationary trajectory, which increases the number of e-folds, and distinctive features on the inflationary observables, 
 such as a reduced tensor to scalar ratio compared with the smooth model,  and a large running of the spectral index. 

\smallskip

For an inflaton $\phi$ with potential $V(\phi)$ in a FRW background, the inflationary evolution is determined by the Friedmann and Klein-Gordon equations:
\bea
&& H^2 = \frac{1}{3M_{Pl}^2 }\left(\frac12 \dot{\phi}^2 + V(\phi)\right) \nn \\
&&  \ddot{\phi} + 3 H \dot{\phi} + V'(\phi) =0 \label{E:Feqs}
\eea
where the Hubble parameter is $H=\dot{a}/a$, $a(t)$ is the scale factor and a dot represents derivatives with respect to cosmic time, while primes represent derivatives with respect to the scalar field. Further $M_{Pl}=(8\pi G)^{-1}$ is the reduced Planck scale, and we use a mostly plus convention for the metric.

As long as $\phi$ is a single-valued function of $t$,  we can express the Hubble parameter as a function of $\phi$, and reformulate these equations into the first order Hamilton-Jacobi (HJ) equations:
\bea
&& H'(\phi)= -\frac{\dot{\phi}}{2 M_{Pl}^2} \nn \\
&& H'(\phi)^2 = \frac{3}{2M_{Pl}^2} H(\phi)^2 - \frac{1}{2M_{Pl}^4} V(\phi) \,. \label{E:HJeqs}
\eea
This formalism allows one to compute the evolution in terms of a new time-variable, the scalar field, $\phi$. 
The cosmological evolution can be characterised via the epsilon parameter
\be
\epsilon  \equiv -\frac{\dot{H}}{H^2} \,.
\ee  
In order for accelerated expansion to occur, that is  $\ddot{a}/a > 0$, this parameter needs to be smaller than unity, $\epsilon<1$.  
In the Hamilton-Jacobi formalism above, one can define a set of HJ slow-roll parameters as follows\footnote{These  parameters are related to the potential slow-roll parameters (\ref{E:potSR}) by $\epsilon \approx \epsilon_V$ and $-4 \delta + 2 \epsilon \approx 2 \eta_V - 4 \epsilon_V$, where the relation ``$\approx$'' holds up to the slow-roll approximation.  Note that when the slow-roll approximation (\ref{E:potSR})  holds, the HJ slow-roll parameters are also small.}: 
\bea\label{HJSR}
&& \epsilon  = 2 M_{Pl}^2 \frac{H'^2}{H^2}\,, \qquad \delta  \equiv M_{Pl}^2 \frac{H''}{H}\,, \qquad \xi  \equiv M_{Pl}^4\frac{H'''H'}{H^2}\,, \qquad \sigma \equiv M_{Pl}^6 \frac{H'''' H'^2}{H^3}\,,  \dots \nn \\ 
\eea
where recall that a prime denotes derivative with respect to the scalar field.  By definition, inflation occurs for as long as the first slow-roll parameter $\epsilon < 1$, and ends when $\epsilon \sim 1$.  Usually, a sufficient period of inflation requires moreover that the second order parameter $\delta < 1$, but as we will see, this is not always necessary.

In order to compute inflationary observables using the slow-roll approximation, the relevant slow variation parameters should   be small around the time of horizon crossing when the observable perturbations were generated. Being in single field inflation, curvature perturbation is then conserved after horizon crossing.   In terms of the slow-roll parameters   \eqref{HJSR}, the observables describing the power spectra of curvature and tensor perturbations are given by (see e.g.~\cite{LythLiddle}):
\bea
{\cal P}_s&=&\left( \frac{H}{4\pi}\right)^2\left(\frac{H}{H'}\right)^2\Bigg |_{\phi = \phi*}
\\
n_s&=& 1 -4\,\epsilon +4 \delta
\\
r&=&16\,\epsilon
\\
\alpha_s&\equiv&\left(\frac{\dif\,n_s}{\dif\,\ln k}\right)_{k=k_*}\,=\,-8\,\xi+20\,\epsilon\,\delta-8\,\epsilon^2
\\
\beta_s&\equiv&\left(\frac{\dif\,\alpha_s}{\dif\,\ln k}\right)_{k=k_*}\,=\,-32\, \epsilon^3+124 \,\epsilon^2 \delta - 80\,
\epsilon\,\delta^2+16 \sigma-56\,  \epsilon \,\xi+16  \,\delta \,\xi \,,
\nonumber\\
\eea
where all parameters are evaluated at horizon crossing determined by the pivot scale $k=k_*$, where also $\phi=\phi_*$.  
 Depending on the physics of reheating and entropy generation, horizon crossing for the pivot scale usually occurs around 50-60 e-folds before the end of inflation.  The   epoch of horizon crossing for the observable  scales probed by the CMB lasts about 7-10 e-folds around this time.

\subsection{Bumpy Monomial Inflation}

We now consider a simple model of a bumpy monomial inflation, which will allow us describe the main features and effects due to oscillations in the smooth potential.  The potential we study is the following\footnote{We use the parameter $A$ to ensure that the potential is vanishing at its minimum.}
\be
V(\phi) = A + \frac12 m^2 \phi^2 + \lambda \phi \cos\left(\frac{\phi}{f}\right) \,.\label{E:ChaotPot}
\ee
These types of potentials are known to arise in supergravity \cite{Kallosh:2014vja} and string theory  (see e.g. \cite{AM3, Tatsuo}) and hence our analysis might be applied to concrete string constructions. When $\lambda=A=0$, we recover $\phi^2$ inflation, which gives a field range of $\Delta \phi \sim 15 \, M_{Pl}$, with a tensor to scalar ratio $r\sim 0.12$  and therefore an inflationary scale of around the GUT scale. While it gives a consistent value for the spectral index $n_s \sim 0.966$, this model is basically excluded by the latest results on $r$.  We now see how this can change when we take into account subleading modulation effects due to higher order corrections such as non-perturbative terms in string theory constructions.  

The physics of the potential \eqref{E:ChaotPot} turns out to be qualitatively different depending on whether or not the oscillations parameterised by $\lambda$ and $f$ introduce new stationary points into the $\phi^2$ potential.  Stationary points are given by the solutions to:
\be
c - \sin x = -\frac{\cos x}{x}\,, \quad \textrm{ where }\quad  c =\frac{ m^2 f}{\lambda} \quad  \textrm{ and } \quad x=\frac{\phi}{f} \,.
\ee
If $ \frac{\lambda}{f} > m^2$, there is an infinite number of stationary points, and classically a rolling scalar field will stop in some local minimum,  depending on its initial conditions \cite{Banks}.  With sufficient initial velocity the endpoint could be the global minimum.  Otherwise, such a potential could provide a background for the ``chain inflation'' \cite{chaininflation} realisation of old inflation, where inflation proceeds through the successive tunneling of the inflaton between local minimum down to the global minimum.  

\begin{figure}[!htb!]
\begin{center}
\includegraphics[width=0.47\textwidth]{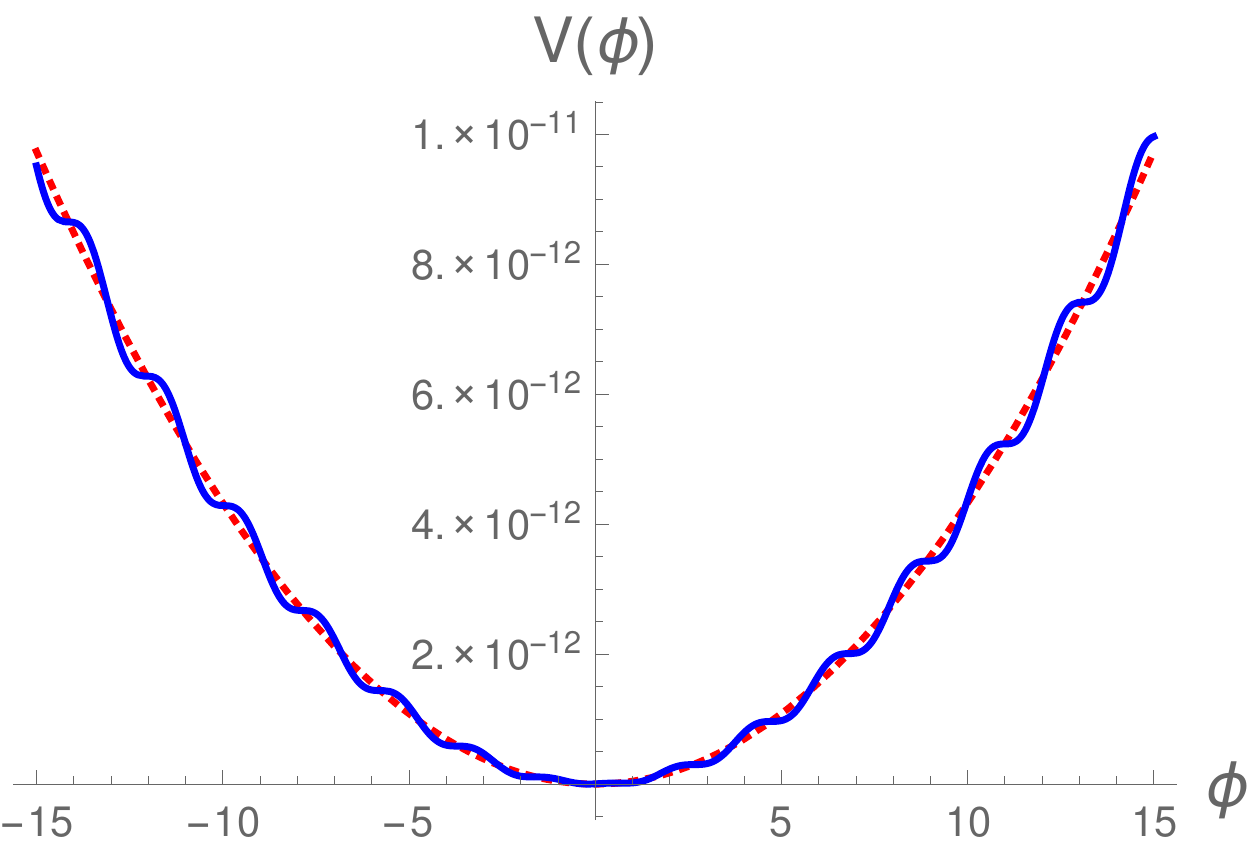}
 \caption{{The bumpy $\phi^2$ potential (\ref{E:ChaotPot}) with $m^2/d^4=10\,M_{Pl}^2$, $\lambda/d^4=3.3\,M_{Pl}^3$, $f=1/3 \,M_{Pl}$, $A/d^4=0.35\,M_{Pl}^4$ (blue, solid), along with the smooth $\phi^2$ potential with $m^2/d^4=10\,M_{Pl}^2$, $\lambda=0$ and $A=0$ (red, dotted), where $d^2 = 9.3 \times 10^{-8}$.  Figure \ref{F:Potential} plots the same potentials for the field range $\phi \in [-5\,M_{Pl},5\,M_{Pl}]$.}}\label{F:ChaotPot}
\end{center}
\end{figure}

Our focus will instead be on smaller (but not much smaller) 
values of $\frac{\lambda}{m^2 f}$, such that no new stationary points are introduced, and  inflation is realised as the scalar field rolls down its bumpy potential, and settles at the minimum.  As an example, take
 $m^2/d^4=10 \, M_{Pl}^2$,  $f=1/3 \, M_{Pl}$, and tune\footnote{For the given $m, f$ and a field range $\Delta \phi \sim 3 \,M_{Pl}$ during inflation, we will need $\lambda/d^4=3.3\, M_{Pl}^3$ to obtain a sufficient number of e-folds.  Decreasing $f$ increases the level of fine-tuning needed in $\lambda$ and widens the necessary field range.  The small modulation models in the literature \cite{AM3, Kobayashi:2010pz, Tatsuo, Kappl:2015esy,Choi:2015aem} have considered smaller axion decay constants, so more frequent modulations.  E.g. axion monodromy models take $f \sim 10^{-2} - 10^{-6} M_{Pl}$, which leads to resonances in the perturbations and oscillations in $n_s$ \cite{AM3}. Interestingly, axions in string theory tend to have large decay constants between around the GUT scale and reduced Planck mass \cite{SW}.} $\lambda/d^4=3.3\, M_{Pl}^3$ to ensure that the turning points in the bumps are very close to stationary points (where we have scaled the parameters by $d^2=9.3\times10^{-8}$ to match the normalisation of scale perturbations (see below) and for convenience with the numerics, also $t\rightarrow t/d^2$).  We draw the potential in Fig. \ref{F:ChaotPot}, together with the corresponding smooth $\phi^2$ model with $\lambda=0$.  Note that several steps are introduced into the potential, each one including a very flat region; our inflationary trajectories will experience a 
 few of these steps. 

\begin{figure}[!htb!]
\centering
\begin{minipage}{.45\textwidth}
\centering
\includegraphics[width=0.9\linewidth]{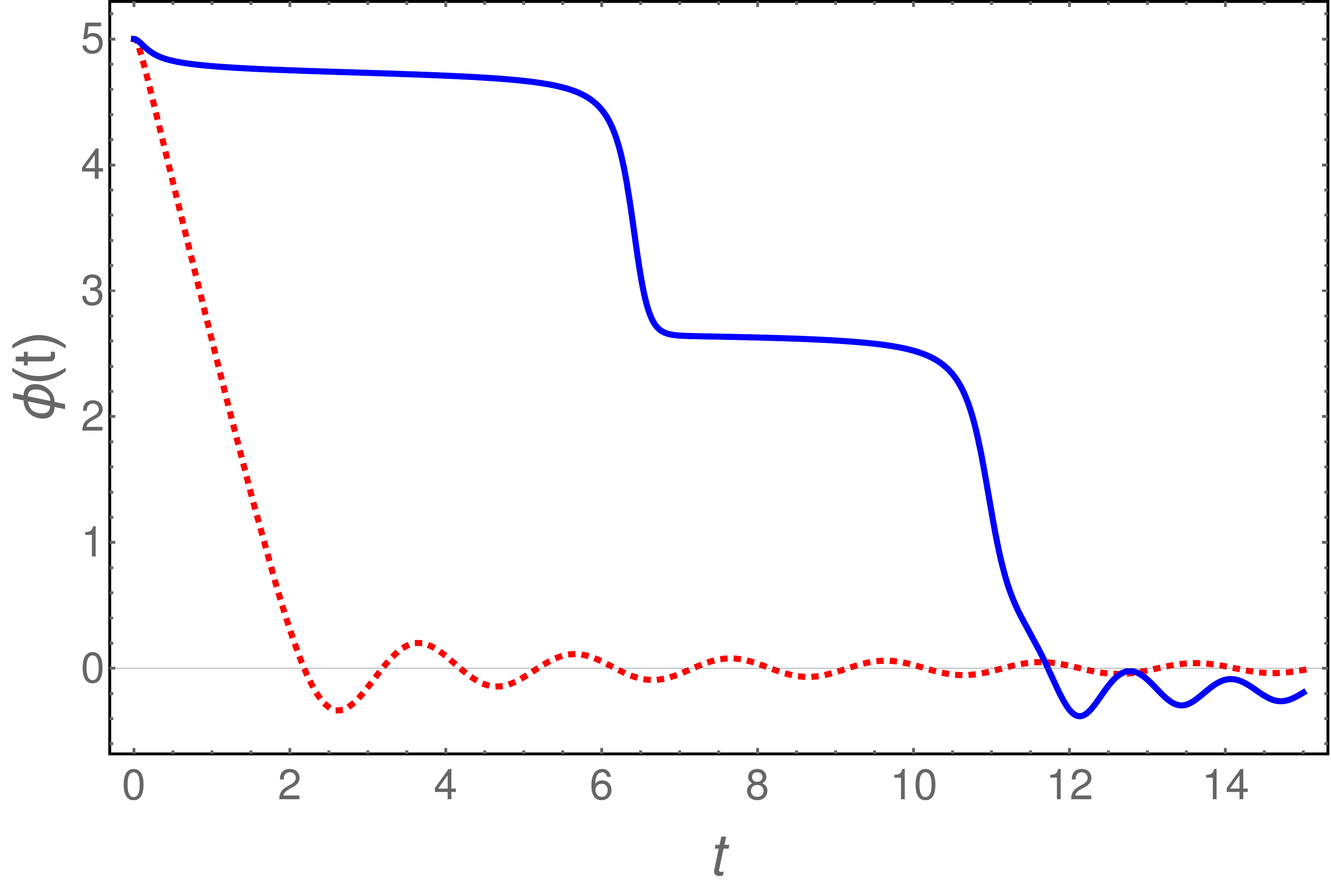}
\end{minipage}
\begin{minipage}{.45\textwidth}
\centering
\includegraphics[width=1.\linewidth]{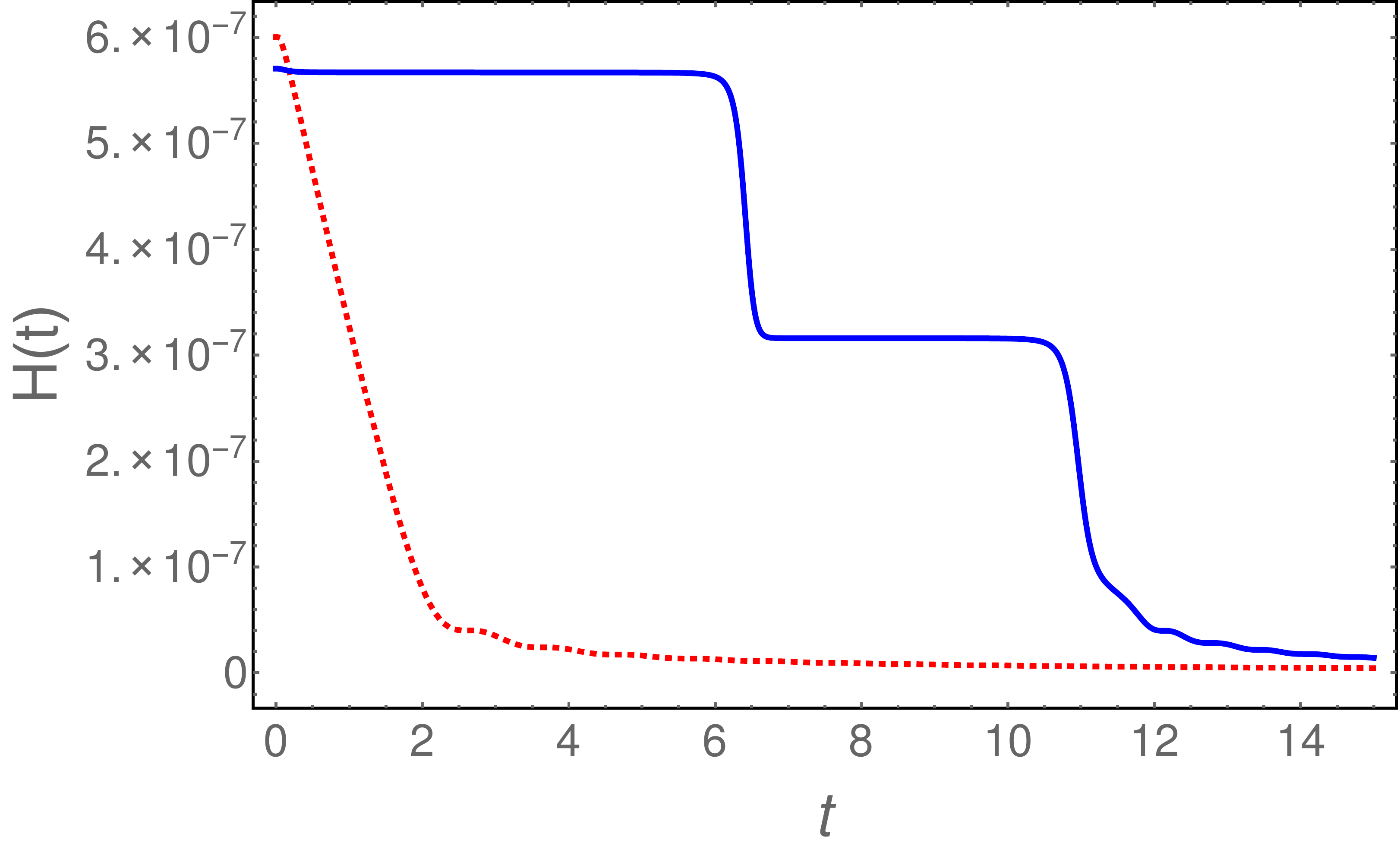}
\end{minipage}
 \caption{{Representation of the solution to the Friedmann equations with the bumpy $\phi^2$ potential (\ref{F:ChaotPot}) (blue, solid) and the smooth $\phi^2$ potential  (\ref{F:ChaotPot}) (red, dotted), for initial conditions $\phi(0)=5 \,M_{Pl}$,  $\dot{\phi}(0)=0$ and $a(0)=1$ and time units $(d^2 M_{Pl})^{-1}$.}}\label{F:ChaotSol}
\end{figure}

For a given choice of initial conditions it is straightforward to solve the Friedmann and Klein-Gordon equations (\ref{E:Feqs}) or Hamilton-Jacobi equations (\ref{E:HJeqs}) numerically.  For example, taking $\phi(0)=5 \,M_{Pl}$, $\dot{\phi}(0)=0$ and $a(0)=1$, we plot the solutions to (\ref{E:Feqs}) along with those of the smooth model in Fig. \ref{F:ChaotSol}.
Independently of where the inflaton starts, the plateaus in the potential slow down the inflaton's progress down the potential,  and keep the Hubble expansion rate $H$ higher for longer.  Indeed, we note empirically that, whilst the inflaton accelerates quickly down the steep cliffs, the Hubble friction (or drag) towards the end of this fast roll is sufficient to avoid an overshoot problem for a wide range of initial velocities\footnote{Even if the initial velocity is so large as to overshoot the first plateau, since the inflaton will eventually slow down to its terminal velocity, it suffices to start higher up in the potential to achieve slow-roll through the plateaus.}, and the inflaton's roll over the subsequent plateau is slow.    Thus the plateaus provide regions of enhanced slow-roll (see also Fig. \ref{F:ChaotSolN} and discussion  below, as well as the analytic model developed in the appendix).  
 
Inflation ends when $\epsilon\simeq1$.  As mentioned above, the smooth $\phi^2$ model yields sufficient e-folds for field range $\Delta \phi \sim 15 \,M_{Pl}$.  Indeed, numerically integrating the above solutions with $\phi(0)=5\,M_{Pl}$ up to the time that inflation ends, we find only $N_{tot} \sim 7$ for the smooth model, whereas $N_{tot} \sim 54$ for the bumpy model.  Thus we see that, for the same initial conditions, the period of inflation is much prolonged in the bumpy model compared to the smooth one, and the number of e-folds much enhanced.   The field range from the beginning to the end of inflation for the bumpy model is therefore reduced compared to 60 e-fold smooth model and in the solution above is $\Delta \phi = 3.4 M_{Pl}$.  
 Notice that although $\epsilon$ stays smaller than 1 for longer, it has small, localised  peaks throughout the phase of inflation, when the inflaton reaches the steep slopes of the steps, leading to large fluctuations for the slow-roll parameters $\delta$, $\xi$ and $\sigma$ (see Fig. \ref{F:ChaoticHSR} and discussion below).

In order to have the first four HJ slow-roll parameters \eqref{HJSR} small around 55 e-folds before the end of inflation, so as to lead to acceptable  inflationary observables, we increase the level of fine-tuning, choosing $\lambda/d^4=0.331 \, M_{Pl}^3$. With the same initial conditions, the total number of e-folds from the beginning to the end of inflation becomes then\footnote{We thus have to confront the generic conceptual Transplanckian problem of inflation, that if inflation lasts more than 70 e-folds, then all scales observed today were subPlanckian at the beginning of inflation \cite{transplanckian}.} $N_{tot}\sim 80$.  The solutions are plotted with respect to the number of e-folds before the end of inflation in Fig. \ref{F:ChaotSolN}.  Thus we see that almost all the e-folds of inflation occur while the inflaton is slowly rolling down the flat plateaus in the bumps.

\begin{figure}[!htb!] 
     \centering
    \includegraphics[width=0.9\linewidth]{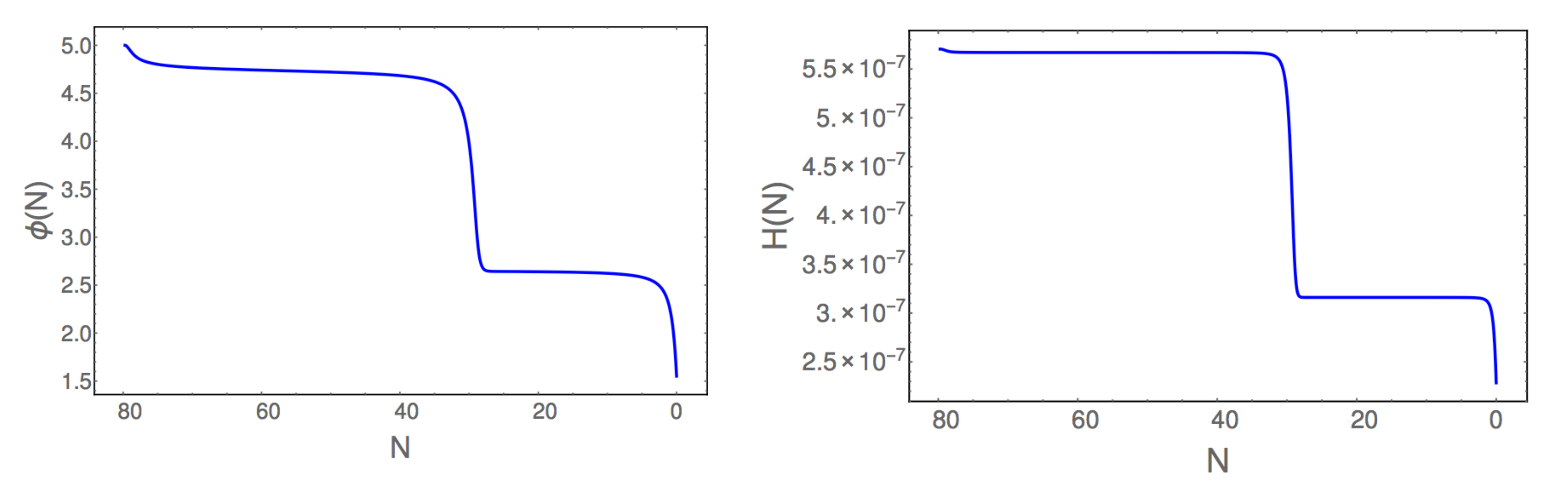} 
    \vspace{4ex}
   \caption{{Solution to the Friedmann equations with the bumpy $\phi^2$ potential (\ref{F:ChaotPot}) with $A/d^4=0.35\,M_{Pl}^4$, $f=1/3 \,M_{Pl}$ and $\lambda/d^4=0.331\,M_{Pl}^3$ and initial conditions $\phi(0)=5\,M_{Pl}$,  $\dot{\phi}(0)=0$ and $a(0)=1$.}}\label{F:ChaotSolN}
  \end{figure}

\begin{figure}[!htb!] 
     \centering
    \includegraphics[width=0.9\linewidth]{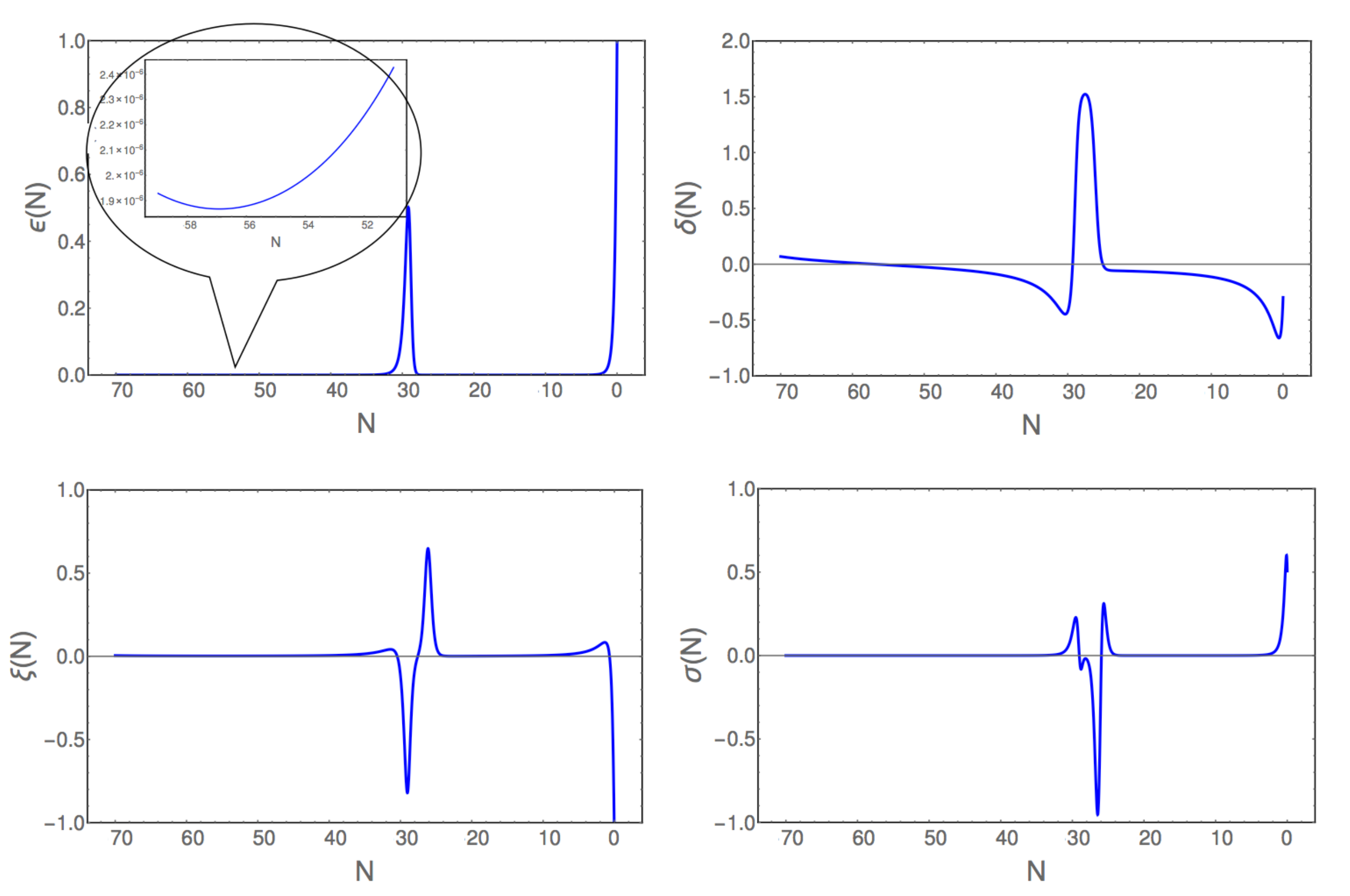} 
    \vspace{4ex}
   \caption{Hubble slow-roll parameters with respect to the number of e-folds before the end of inflation, for the bumpy $\phi^2$ model solution, Fig. \ref{F:ChaotSolN}.}\label{F:ChaoticHSR}
\end{figure}

We also show how the Hubble slow-roll parameters vary with the number of e-folds throughout inflation in Fig. \ref{F:ChaoticHSR}.  During the whole inflationary epoch ($\sim$65 e-folds from the largest scales observable today up to the end of inflation), the Hubble slow-roll parameters undergo strong oscillations, when the inflaton rolls down the steep slopes of the bumps.  However, during the shorter range of e-folds 
which can be probed observationally by the CMB,  ($\sim$10 e-folds around $N=50-60$), all the slow-roll parameters are small and  smoothly varying. 
This implies that we do not expect consequent features in the power spectrum or non-Gaussian observables (like the ones explored for example in \cite{Chen:2008wn,Flauger:2009ab}), since they can occur only at
scales not probed by current CMB observations. 

\begin{figure}[!htb!]
\centering
\begin{minipage}{.45\textwidth}
\centering
\includegraphics[width=0.9\linewidth]{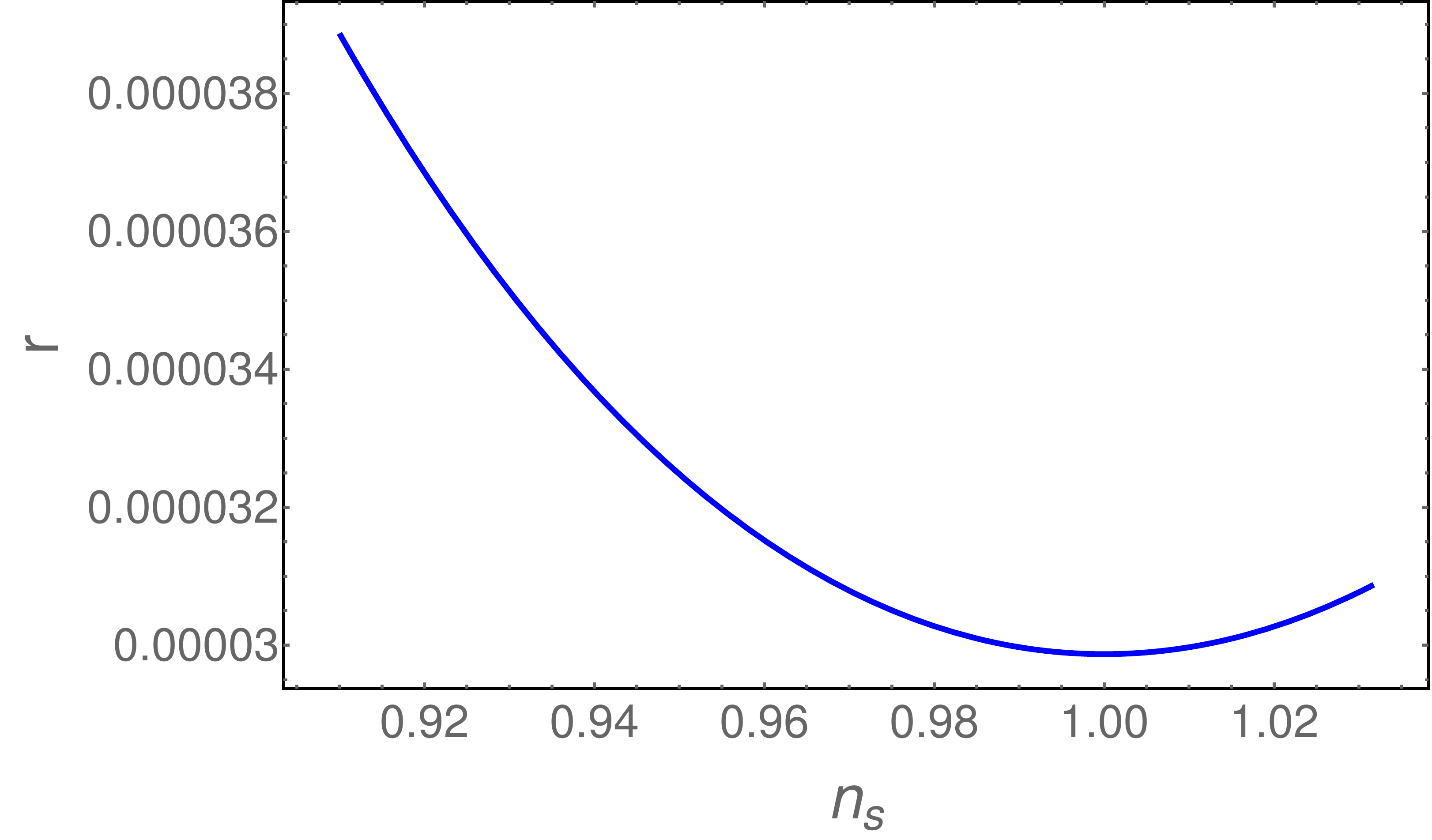}
\end{minipage}
\begin{minipage}{.45\textwidth}
\centering
\includegraphics[width=0.9\linewidth]{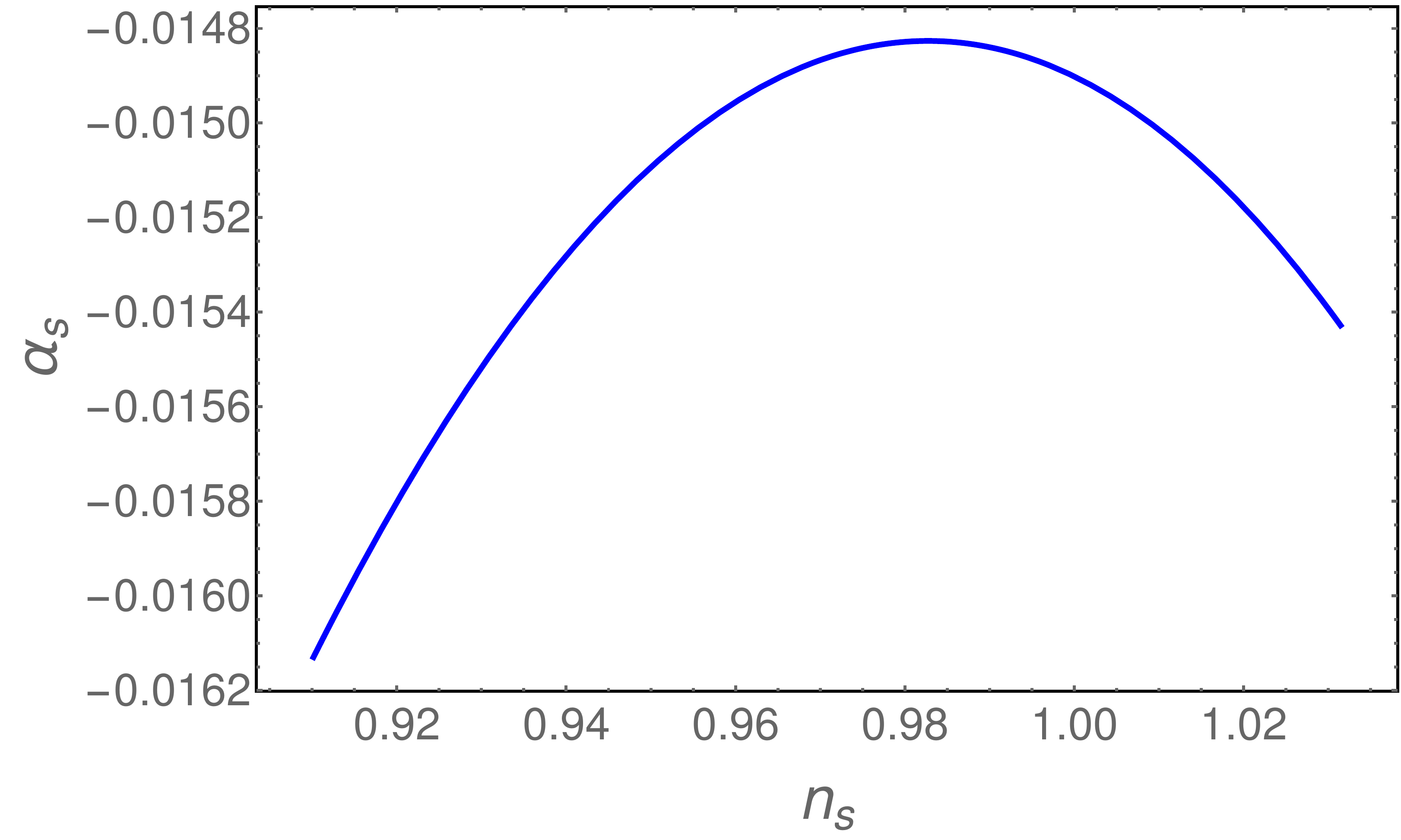}
\end{minipage}
 \caption{{Values for $(n_s,r)$ and $(n_s,\alpha_s)$ for horizon crossing between $N=51$ and $N=59$ e-folds before the end of inflation, for the bumpy $\phi^2$ model solution.}}\label{F:Chaotnsr}
\end{figure}

\begin{figure}[!htb!]
\centering
\begin{minipage}{.45\textwidth}
\centering
\includegraphics[width=0.9\linewidth]{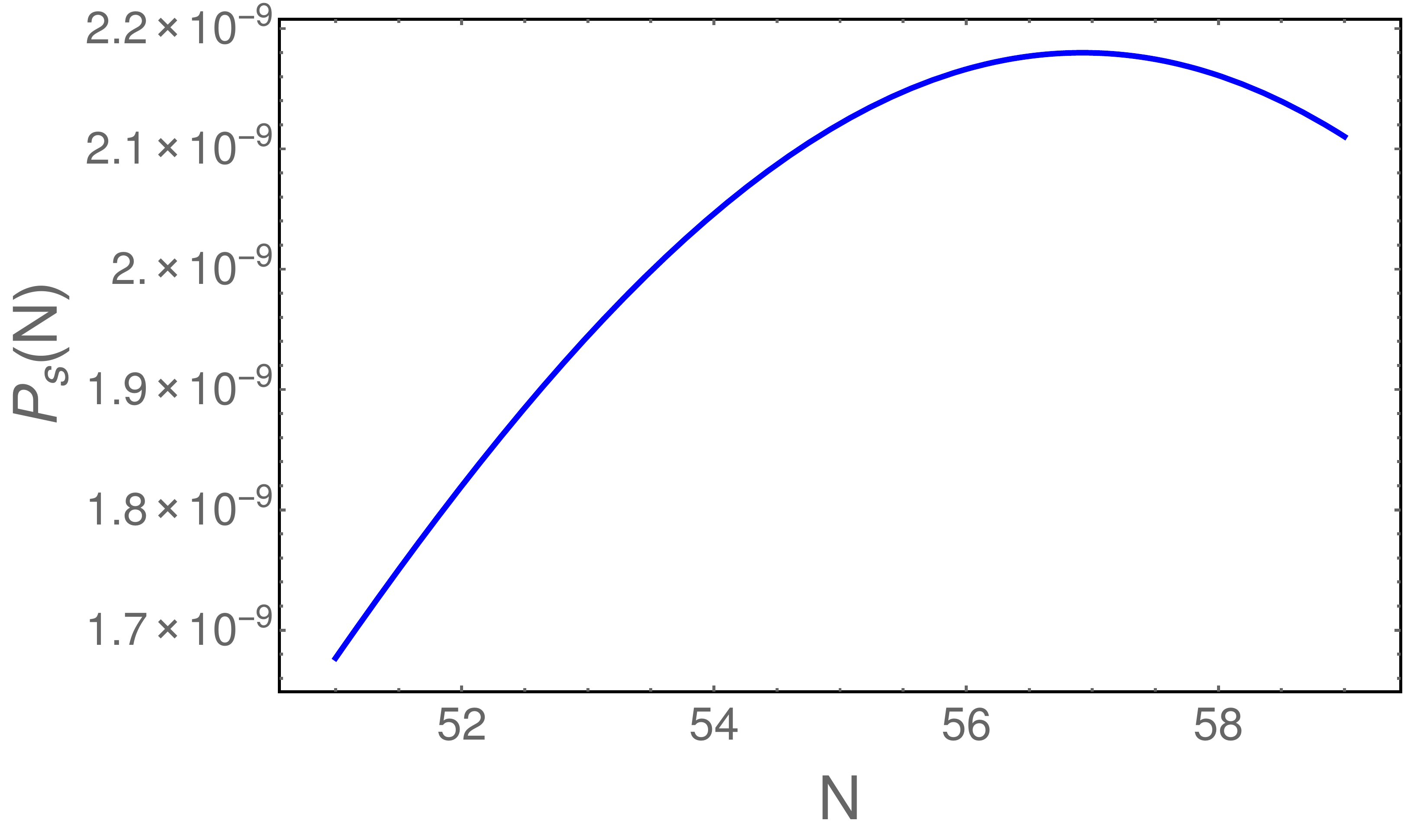}
\end{minipage}
\begin{minipage}{.45\textwidth}
\centering
\includegraphics[width=0.9\linewidth]{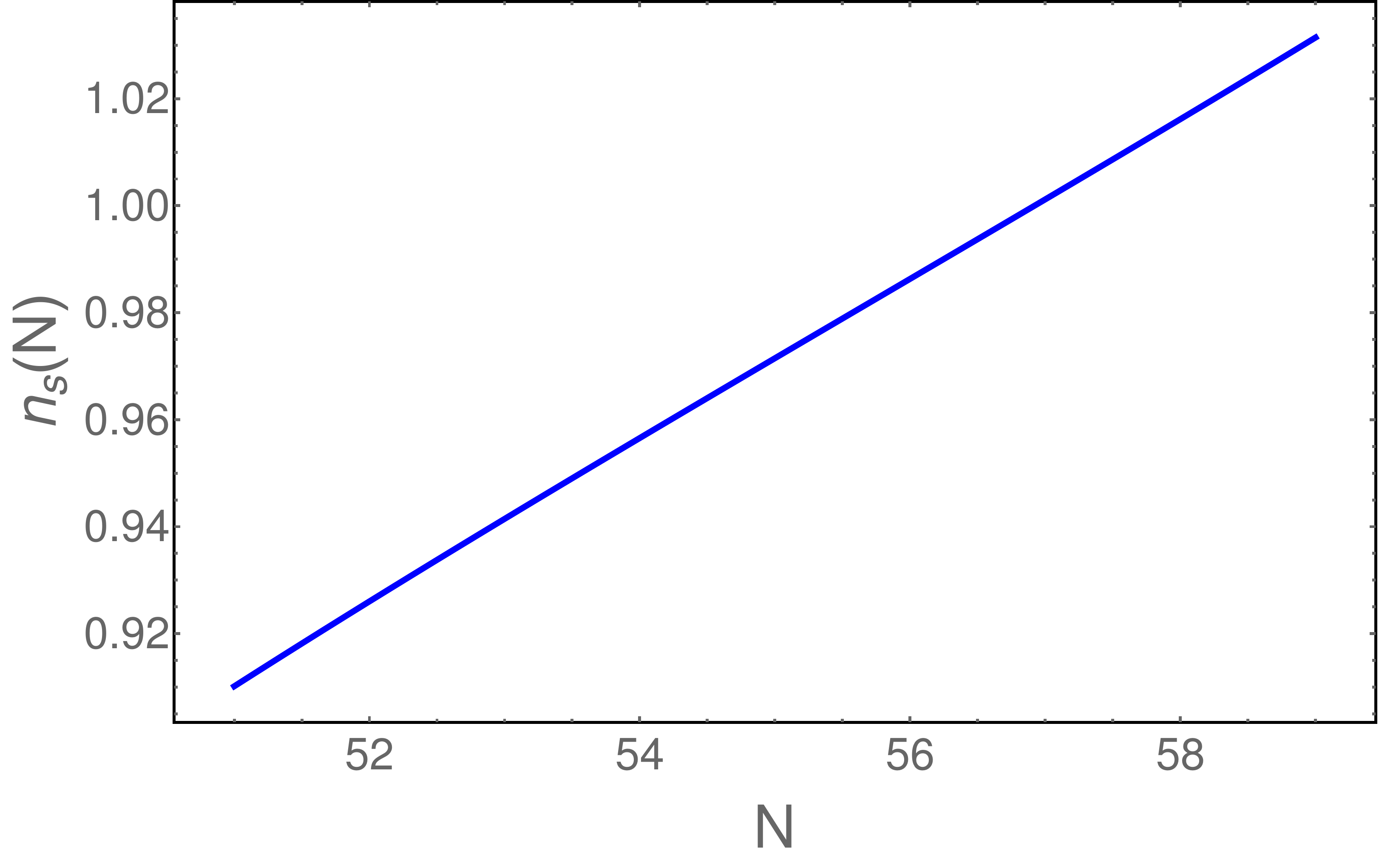}
\end{minipage}
 \caption{{Variation of perturbation amplitude, $P_s$, and spectral index, $n_s$, during the epoch of horizon crossing probed by the CMB, for the bumpy $\phi^2$ model solution.}}\label{F:ChaotnsPs}
\end{figure}

One can now easily compute the CMB observables from the slow-roll parameters at horizon crossing for the pivot scale.  For example, assume the pivot scale crossed the horizon at around $N_e\sim 55$ e-folds before the end of inflation.  As indicated earlier, all the parameters in the scalar potential (and time, $t$) have been scaled to match the observed amplitude for scalar perturbations, $P_s \approx 2.1 \times 10^{-9}$ \cite{Planck15Infla}.  The HJ slow-roll parameters at horizon crossing are:
\bea
&& \epsilon = 1.9 \times 10^{-6}\,, \qquad \delta = -0.0083\,, \qquad \xi = 0.0019 \qquad \textrm{ and } \qquad \sigma = 8.0 \times 10^{-6} \,,
\eea
 yielding the following values for the remaining observables
\bea
&& n_s = 0.9667\,, \qquad r = 3.1 \times 10^{-5}\,, \qquad \alpha_s= -0.015 \qquad \textrm{ and } \qquad \beta_s = -1.2 \times 10^{-4}\,, \label{E:chaoticobs}
\eea
in the ballpark of the Planck measurements and constraints given in (\ref{E:Planck2}).  We plot the tensor to scalar ratio, $r$, and running of the spectral index, $\alpha_s$, against the value of the spectral index, $n_s$, in Fig. \ref{F:Chaotnsr}, for a range of horizon crossings between $N=51$ and $N=59$ e-folds before the end of inflation.
Although the slow-roll parameters vary slowly during the epoch of horizon crossing, the slope in $\delta$ leads to a large running $\alpha_s$ of the spectral index, so that the spectral tilt $n_s$ varies significantly over the scales probed by the CMB.  In Fig. \ref{F:ChaotnsPs}, we plot the variation of the scalar perturbation amplitude and spectral index during the epoch of horizon crossing.  It would be important to understand if such variation could be detected or ruled out in current or future measurements of the CMB \cite{Turner}.

The field range\footnote{Note that the values for $r$ and $\Delta \phi$ are in agreement with the more precise study of the Lyth bound, which takes into account the effect of the tilt as discussed in \cite{GRSZ2}.} for the bumpy model from horizon crossing to the end of inflation is $\Delta\phi \sim 3.2 \, M_{Pl}$, and the scale of inflation at horizon crossing is:
\be
(V_{inf})^{1/4} = 9.9 \times 10^{-4} M_{Pl} \,.
\ee
Thus, non-perturbative corrections can modify a large field monomial model of inflation to one with intermediate field range and inflationary scale, making a consistent perturbative string theory realisation of the model possible.   The distinctive signature of such a scenario is a suppressed tensor mode and a large negative running of the spectral index.

\subsection{Bumpy Natural Inflation}

We now consider a single field modulated natural potential\footnote{Multifield modulated  natural inflation models have been discussed recently in \cite{Kappl:2015esy,Choi:2015aem}.}, which is generically predicted by string theory models. As we will show, similar results can be found as for the single field bumpy monomial potential. A simple model that encapsulates the  physics we are interested in  is: 
\be
V(\phi) = A + \Lambda^4\left(1+\cos\left(\frac{\phi}{f}\right)\right) +  \tilde \Lambda^4\left(1+\cos\left(\frac{\phi}{\tilde f}\right)\right) \label{E:NatPot}
\ee
where $\tilde \Lambda<\Lambda$ and $\tilde f < f$ parameterise the bumps.
Stationary points in the potential are given by the solutions to:
\be 
\sin x = - b \sin c \,x \, \quad \textrm{ where } \quad b = \frac{\tilde \Lambda^4}{\Lambda^4} \frac{f}{\tilde f} \quad \textrm{ and } \quad 
c = \frac{f}{\tilde f},
\ee
where $x$  is defined as before. For example, choosing $\Lambda^4/d^4 = 1 \, M_{Pl}^4 $, $f=1 \, M_{Pl}$ and $\tilde f=1/3 \,M_{Pl}$, we tune\footnote{We need to fine-tune $\tilde \Lambda^4$ to 4 decimal points to obtain  sufficient number of e-folds.  Decreasing $\tilde f$, increases the level of fine-tuning required.} $\tilde \Lambda^4/d^4 = 0.3329\, M_{Pl}^4$ (where now $d^2=9.1\times 10^{-8}$) to ensure the turns in the potential are close to stationary points.  The potential is plotted below, along with that of smooth natural inflation.  The bumpy model again has a step-like shape, with steep regions connected by a plateau.
\begin{figure}[!htb!]
\begin{center}
\includegraphics[width=0.37\textwidth]{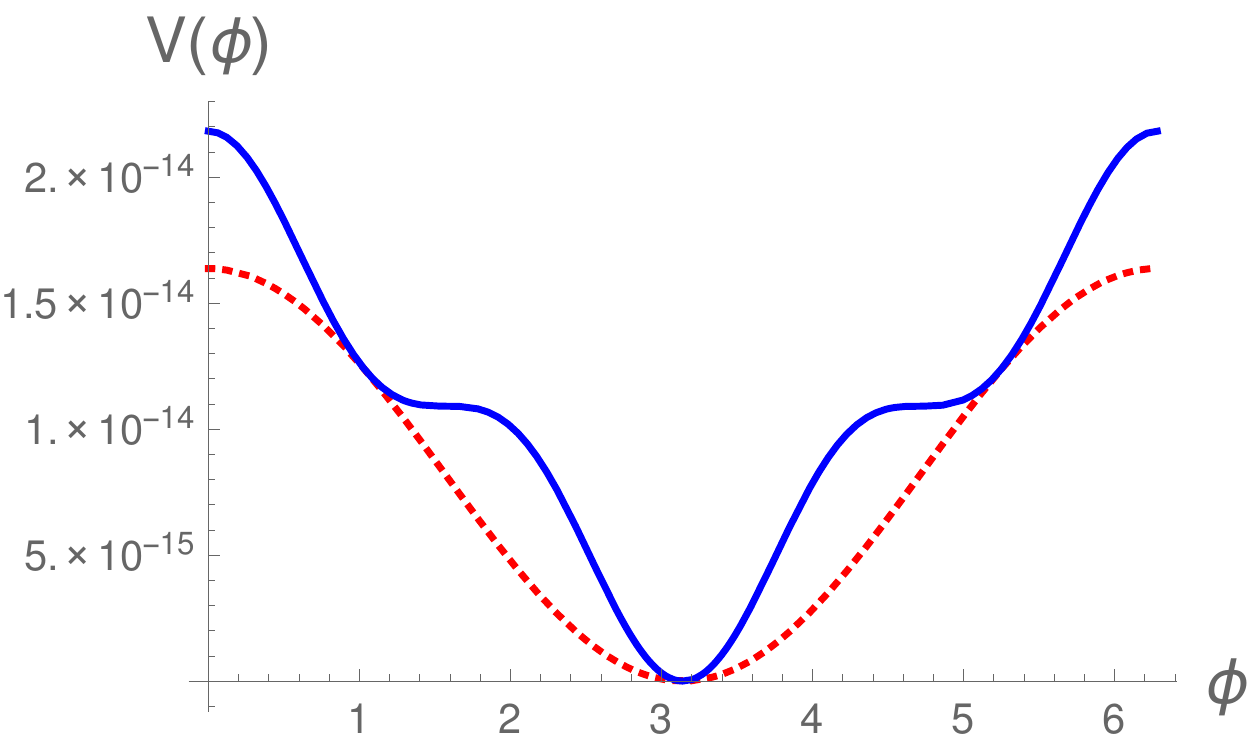}
 \caption{{The bumpy natural potential (\ref{E:NatPot}) with $\Lambda^4/d^4=1\,M_{Pl}^4$, $f=1\, M_{Pl}$, $\tilde \Lambda^4/d^4=0.3329\,M_{Pl}^4$, $f=1/3\, M_{Pl}$, $A=0$ (blue, solid), along with the smooth natural potential with $\Lambda^4/d^4=1\, M_{Pl}^4$, $f=1\, M_{Pl}$, $\tilde \Lambda^4=0$, $A=0$ (red, dotted), where now $d^2=9.1\times 10^{-8}$.}}\label{F:NatPot}
\end{center}
\end{figure}  
The solutions to the corresponding Friedmann equations are given in Fig. \ref{F:NatSol}, for initial conditions  $\phi(0)=0.001\,M_{Pl}$,  $\dot{\phi}(0)=0$ and $a(0)=1$.   Just as in the $\phi^2$ case, for a wide range of initial conditions\footnote{For an initial velocity so high that the inflaton overshoots the plateau, an increase in $f$ and the field range would give the Hubble friction sufficient time to slow down the inflaton to its terminal velocity and prevent overshoot.} the plateau slows down the overall progress of the inflaton to its minimum, and prolongs the phase of inflation defined by $\epsilon <1$.  The number of e-folds achieved during the inflationary phase is $N_{tot} \sim 62$ (compared with $N_{tot}\sim 18$ for the smooth model), and the field range is only $\Delta\phi = 2.6 M_{Pl}$.  Remarkably, a sufficient number of e-folds has been achieved from a single inflaton with Planckian axion decay constant and moderate inflaton field range.
 
\begin{figure}[!htb!]
\centering
\begin{minipage}{.45\textwidth}
\centering
\includegraphics[width=0.9\linewidth]{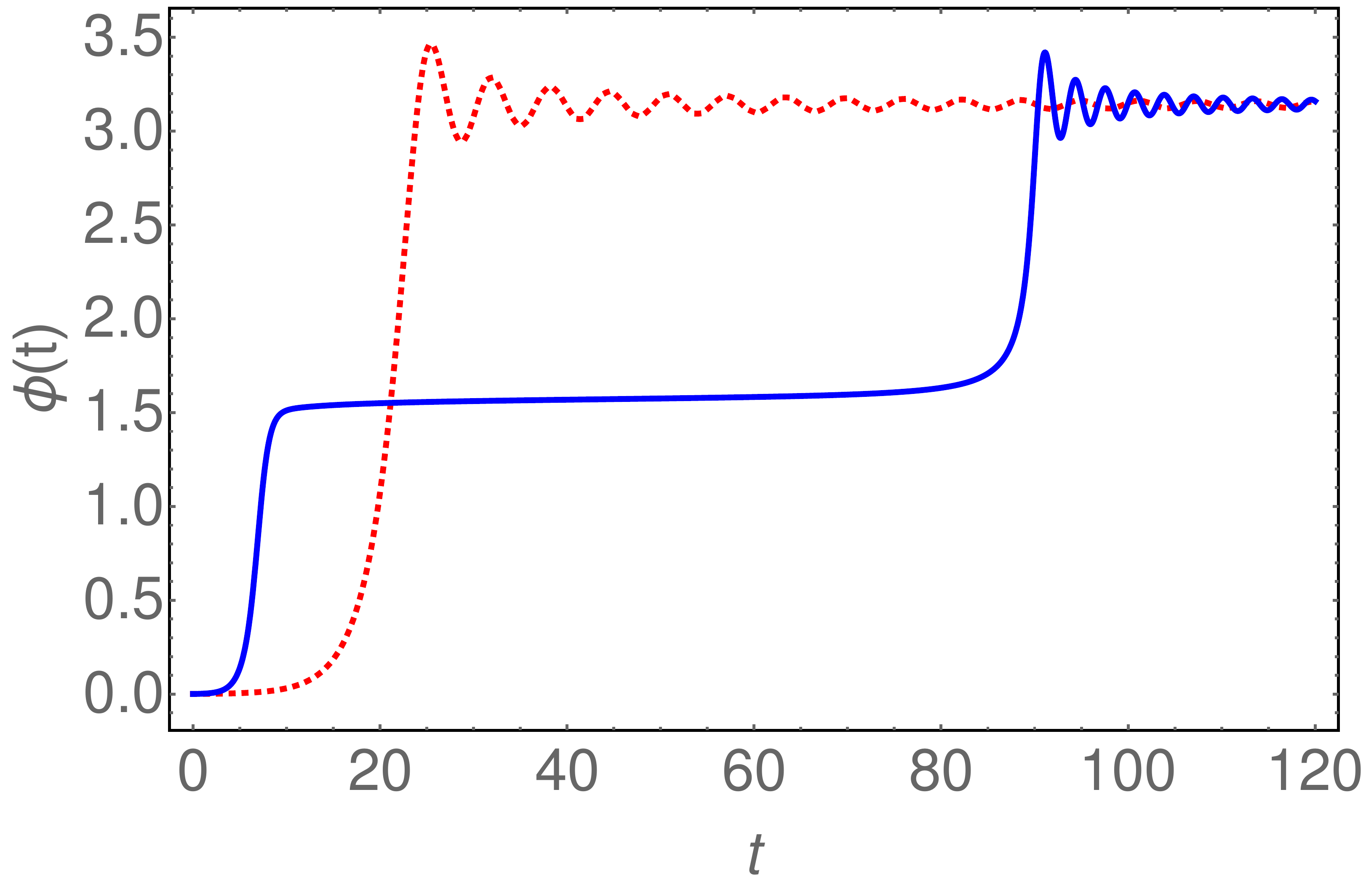}
\end{minipage}
\begin{minipage}{.45\textwidth}
\centering
\includegraphics[width=1.0\linewidth]{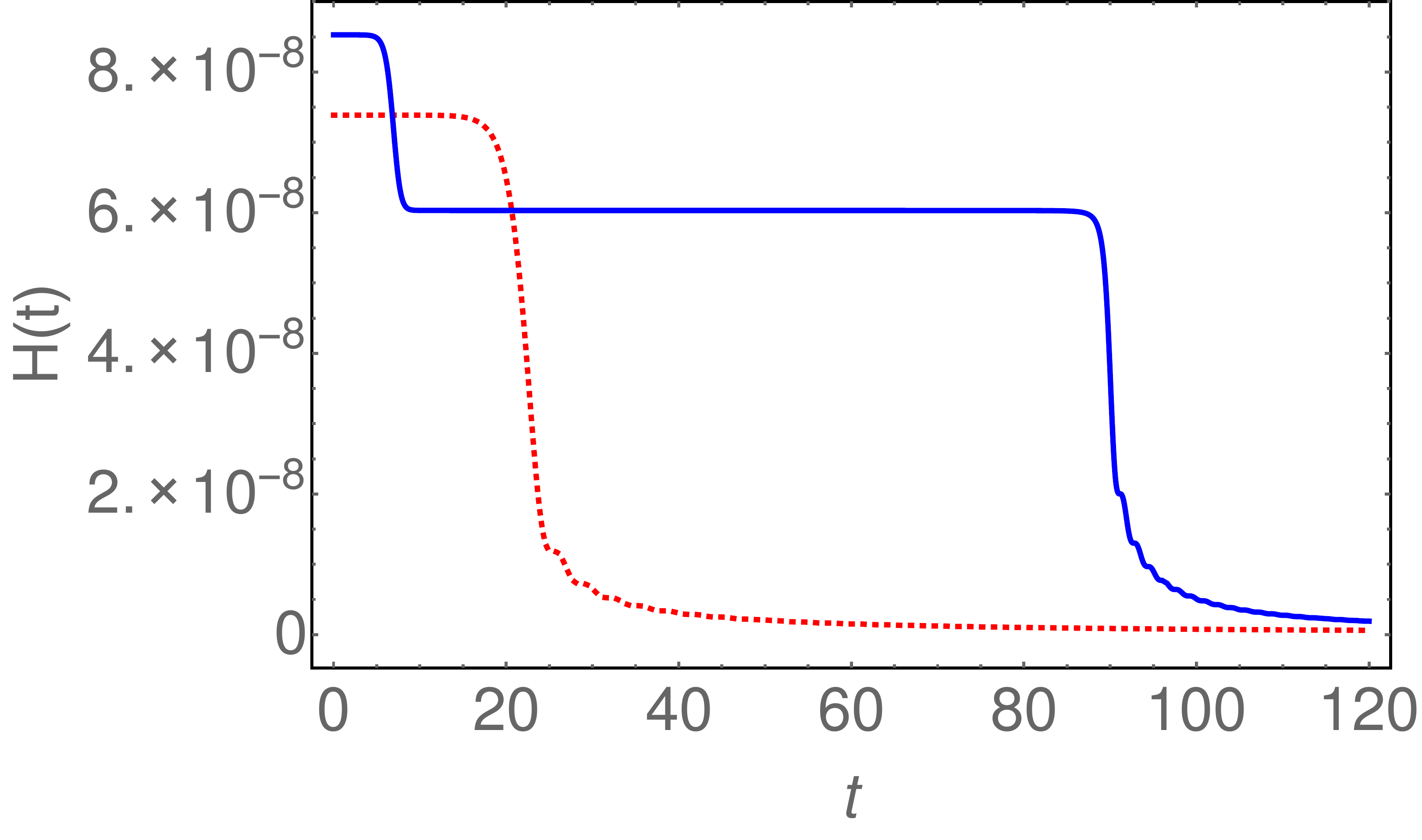}
\end{minipage}
 \caption{{Solution to the Friedmann equations with the bumpy natural potential (\ref{F:NatPot}) (blue, solid) and the smooth natural potential  (\ref{F:NatPot}) (red, dotted), for initial conditions $\phi(0)=0.001\, M_{Pl}$,  $\dot{\phi}(0)=0$ and $a(0)=1$, in units of time $(d^2 M_{Pl})^{-1}$.
 }}\label{F:NatSol}
\end{figure}

\begin{figure}[!htb!] 
     \centering
    \includegraphics[width=0.9\linewidth]{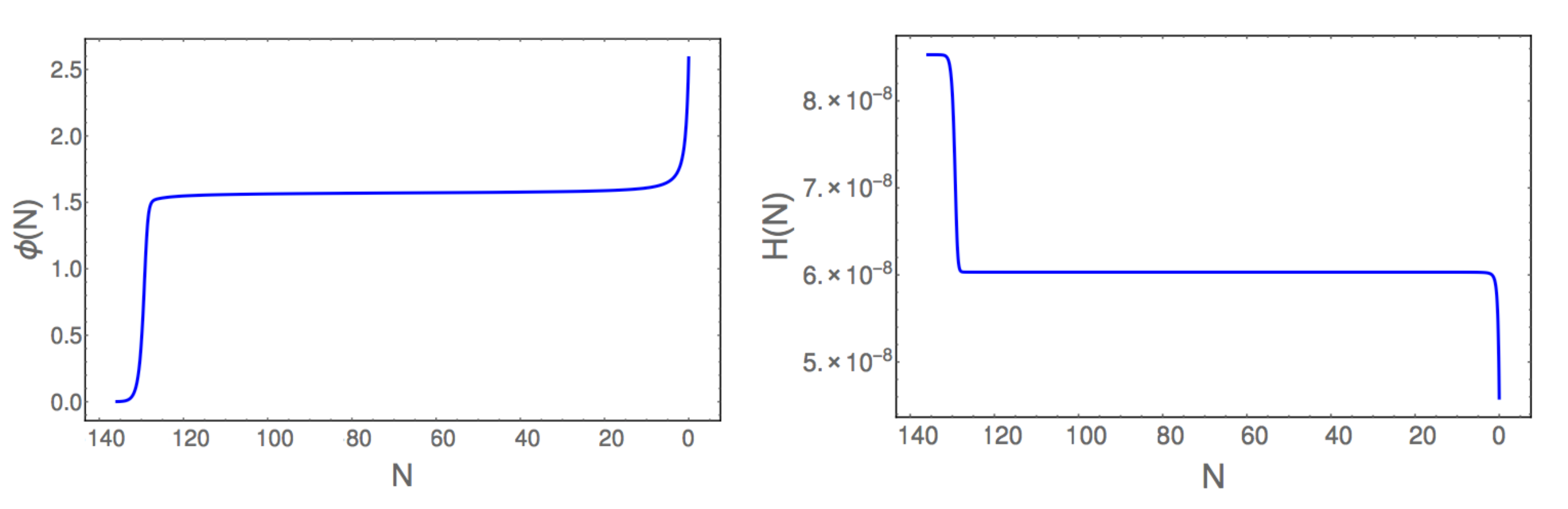} 
    \vspace{4ex}
  \caption{{Solution to the Friedmann equations as functions of $N$ with the bumpy natural potential (\ref{F:NatPot}) with $A=0$, $f=1/3\, M_{Pl}$ and $\lambda/d^4=0.33325\, M_{Pl}^4$ and initial conditions $\phi(0)=0.001\, M_{Pl}$,  $\dot{\phi}(0)=0$ and $a(0)=1$.}}\label{F:NaturalSolN}
\end{figure}

To obtain acceptable values for the inflationary observables, we increase the level of fine-tuning, taking $\tilde \Lambda^4/d^4=0.33325\, M_{Pl}^4$.  For initial conditions we use again $\phi(0)=0.001\, M_{Pl}$, $\dot{\phi}(0)=0$ and $a(0)=1$.  The total number of e-folds from the beginning to the end of inflation is $N_{tot}\sim 136$.  By plotting the inflaton and Hubble parameter with respect to the number of e-folds before the end of inflation in Fig. \ref{F:NaturalSolN}, we see that almost all the inflation occurs while the inflaton is on the plateau.

  \begin{figure}[!htb!] 
     \centering
    \includegraphics[width=0.9\linewidth]{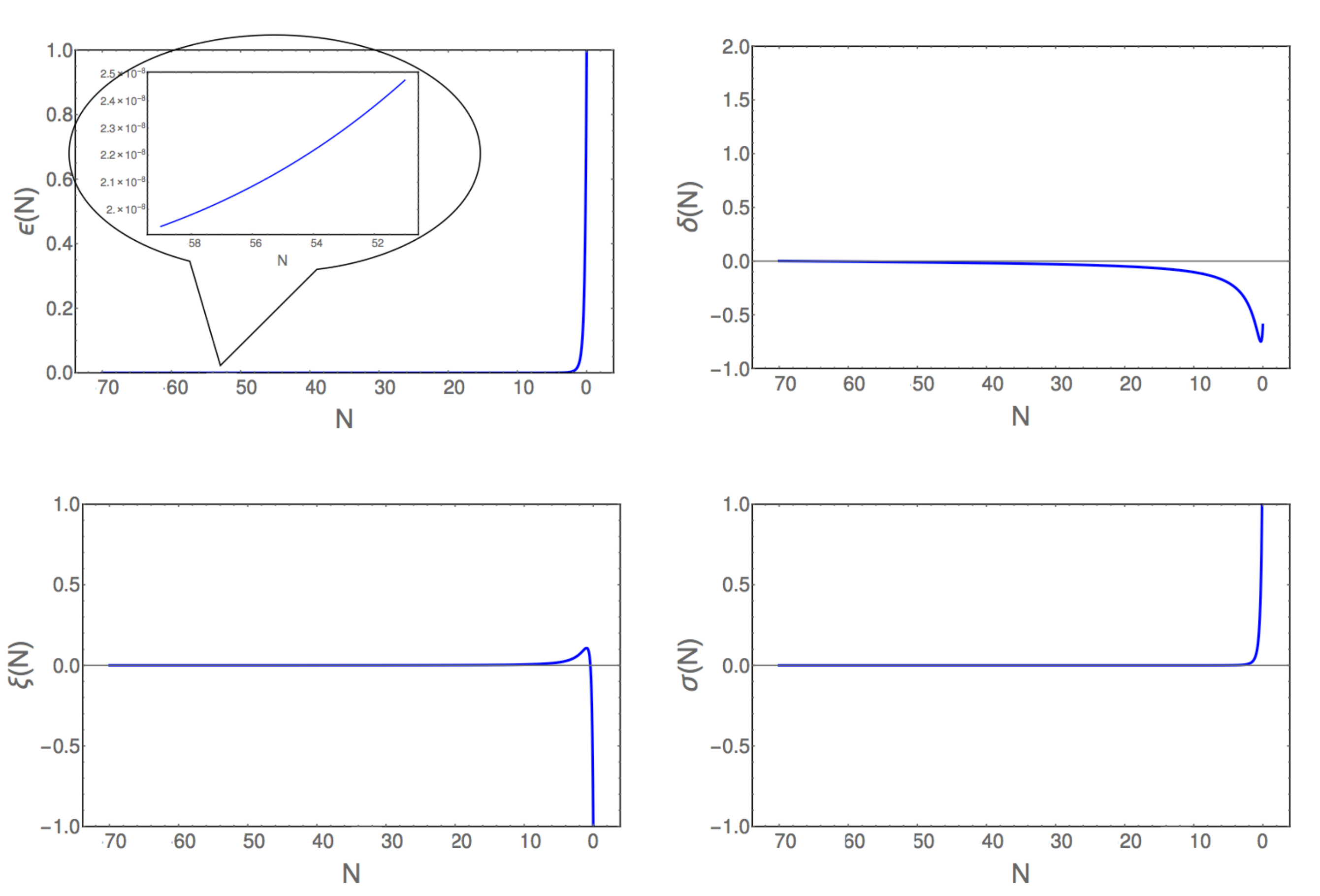} 
    \vspace{4ex}
 \caption{Slow-roll parameters \eqref{HJSR} with respect to the number of e-folds before the end of inflation, for the bumpy natural solution, Fig. \ref{F:NaturalSolN}.}\label{F:NatHSR}
\end{figure}

\begin{figure}[!htb!]
\centering
\begin{minipage}{.45\textwidth}
\centering
\includegraphics[width=0.9\linewidth]{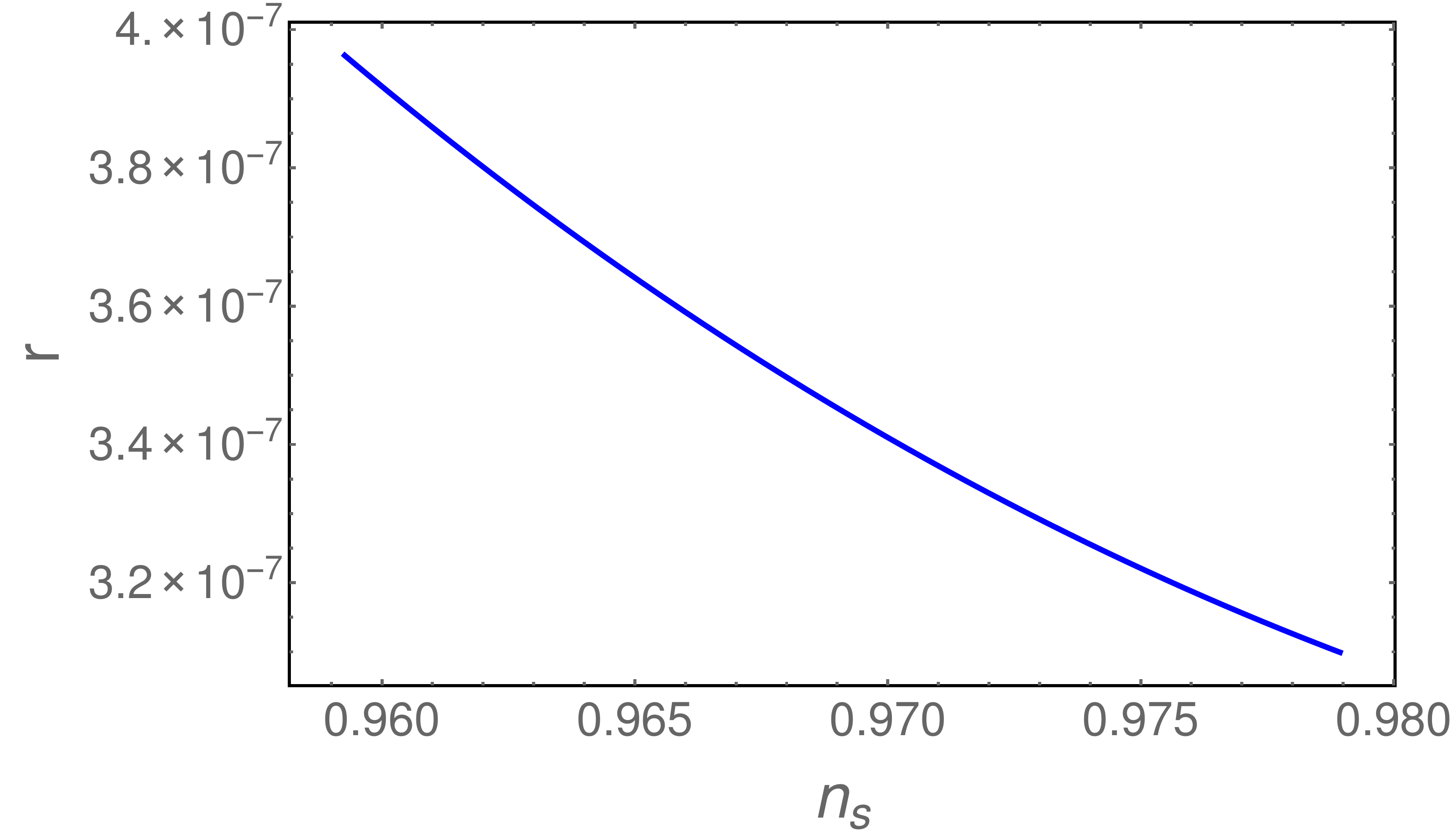}
\end{minipage}
\begin{minipage}{.45\textwidth}
\centering
\includegraphics[width=0.9\linewidth]{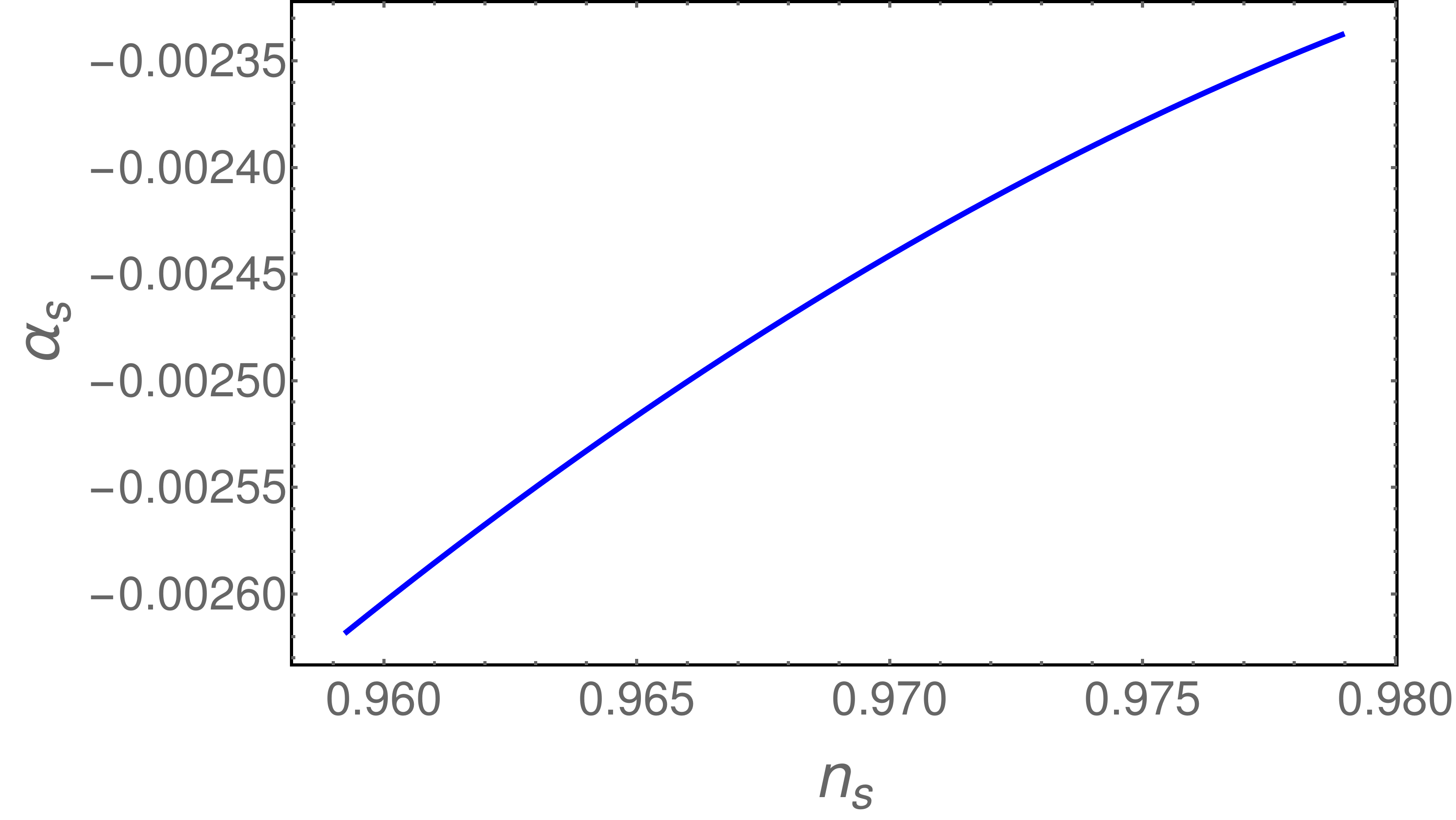}
\end{minipage}
 \caption{{Values of $(n_s,r)$ and $(n_s,\alpha_s)$, for horizon crossing between $N=51$ and $N=59$ e-folds before the end of inflation.}}\label{F:Naturalnsr}
\end{figure}

\begin{figure}[!htb!]
\centering
\begin{minipage}{.45\textwidth}
\centering
\includegraphics[width=0.9\linewidth]{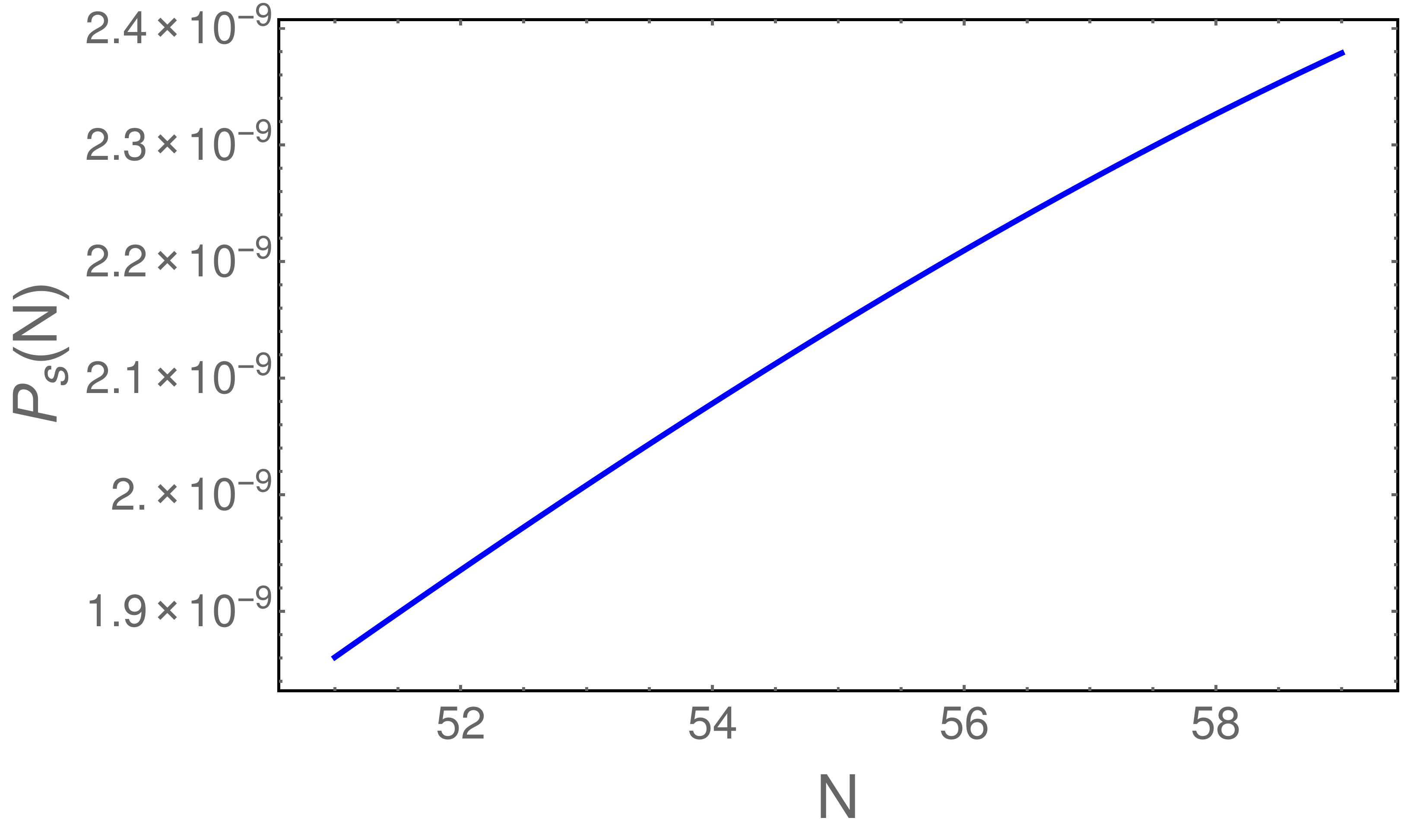}
\end{minipage}
\begin{minipage}{.45\textwidth}
\centering
\includegraphics[width=0.9\linewidth]{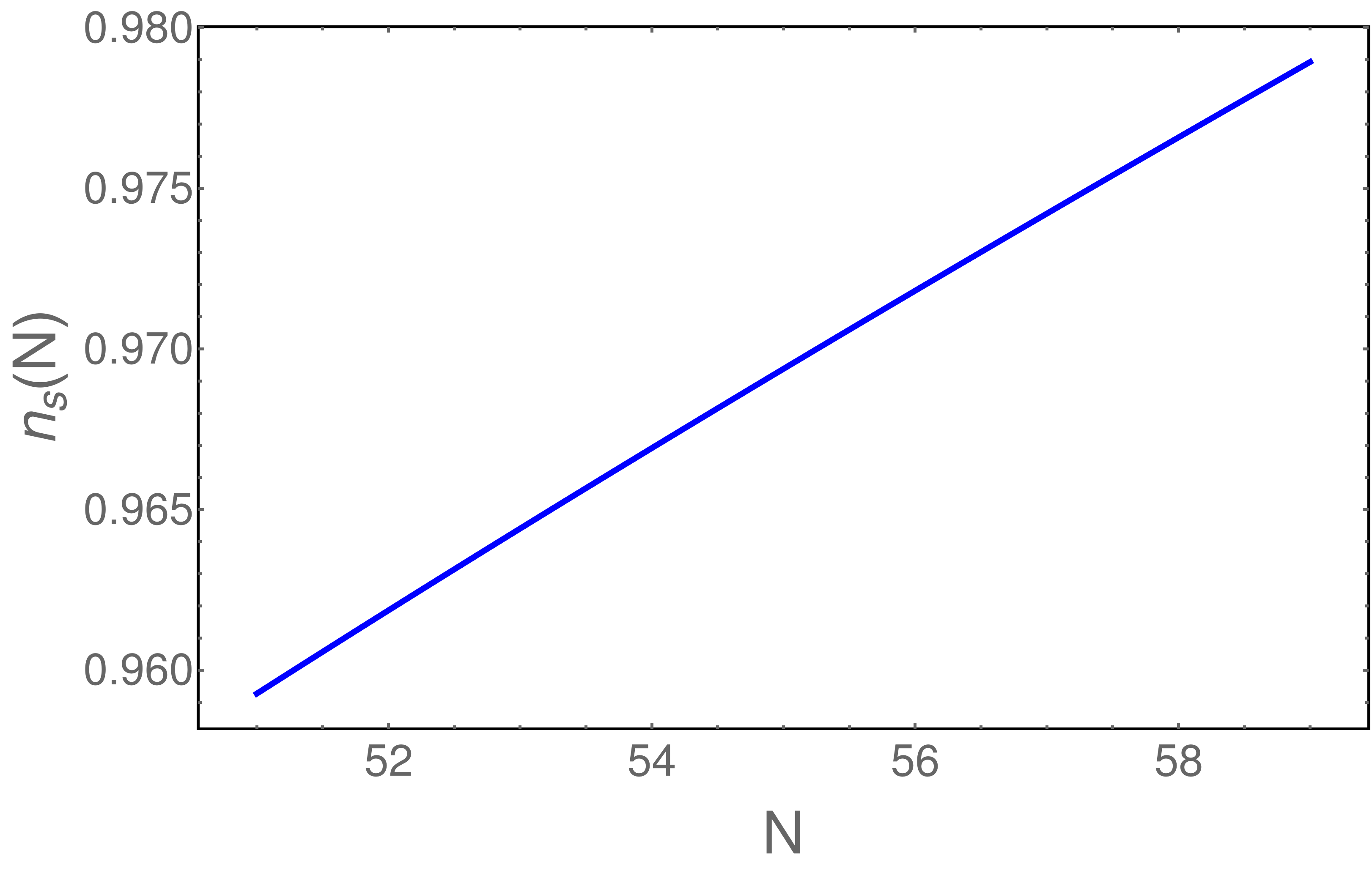}
\end{minipage}
 \caption{{Variation of perturbation amplitude, $P_s$, and spectral index, $n_s$, during the epoch of horizon crossing probed by the CMB, for the bumpy natural solution.}}\label{F:NatPsns}
\end{figure}

The Hubble slow-roll parameters are given in Fig. \ref{F:NatHSR}. Almost all the inflation proceeds in a slow-roll fashion, although there are large fluctuations in the slow-roll parameters when rolling down the steep slopes of the steps.  In particular, as the Hubble slow-roll parameters are all small during horizon crossing epoch around $N_e=50-60$ e-folds before the end of inflation, we can use the slow-roll approximation to compute the CMB observables.  Take the pivot scale to cross the horizon at $N_e\sim 54$ e-folds before the end of inflation.  As already mentioned, the potential (and time, $t$) has been scaled to match the observed amplitude for scalar perturbations, $P_s \approx 2.1 \times 10^{-9}$.  The Hubble slow-roll parameters at horizon crossing are:
\bea
&& \epsilon = 2.2 \times 10^{-8}\,, \quad \delta = -0.0081\,, \quad \xi= -0.00031 \quad \textrm{ and } \quad \sigma = 1.9 \times 10^{-7} \,,
\eea
 yielding the following values for the remaining observables:
\bea
&& n_s = 0.9677\,, \quad r = 3.5 \times 10^{-7}\,, \quad \alpha_s=-0.0025 \quad \textrm{ and } \quad \beta_s= -3.7 \times 10^{-5} \,,
\eea
in agreement with the Planck measurements given in (\ref{E:Planck1}).  As in the monomial model, tensor modes are undetectably small, whereas there is a large negative running of the spectral index (one order of magnitude smaller than that in the monomial case).  Allowing for a range of e-folds at horizon crossing between $N=51$ and $N=59$, we plot the tensor to scalar ratio, $r$, and the running of the spectral index, $\alpha_s$, against the spectral index, $n_s$, in Fig. \ref{F:Naturalnsr}.
The variation of $P_s$ and $n_s$ during the full epoch of horizon crossing probed by the CMB is plotted in Fig. \ref{F:NatPsns}.  Again, it would be important to understand if such a variation could be detected or ruled out in the CMB. 

The field range from horizon crossing to the end of inflation is $\Delta \phi = 1.0 \, M_{Pl}$ and the scale of inflation at horizon crossing is:
\be
(V_{inf})^{1/4} = 3.2 \times 10^{-4} M_{Pl} \,.
\ee
We reiterate that non-perturbative corrections have made it possible to achieve single field, natural inflation with a Planckian axion decay constant, $f \sim M_{Pl}$, moderate field range and intermediate inflationary energy scale.  Thus a consistent embedding in perturbative string theory becomes possible.

\section{Discussion}
\label{S:Conclusions}  

Axions provide a very interesting class of inflationary models which are well-motivated from theories of quantum gravity like string theory, and  are protected from dangerous perturbative corrections to the effective potential.  Non-perturbative corrections typically introduce modulations into monomial and natural inflaton potentials.  Whereas it was previously assumed that such bumps would either spoil slow-roll inflation \cite{Banks} or produce negligible corrections to the slow-roll dynamics \cite{AM1, Kobayashi:2010pz, Tatsuo, Kappl:2015esy,Choi:2015aem}, we have seen that these effects can lead to an enhancement of the number of e-folds.  In particular, when the bumps take the form a series of steep cliffs and gentle plateaus, the large Hubble friction (or drag) and the sharp reduction in acceleration at the end of the cliffs, cause the inflaton to roll slowly whenever it reaches a plateau.  Hence, although slow-roll parameters have large fluctuations through the sharp cliffs, the plateaus provide regions of slow-roll that produce many e-folds of inflation. Our scenarios are thus single field models of inflation, in which slow-roll conditions are not always satisfied during the inflationary trajectory.

Consequently, both single field monomial and natural inflation models can give sufficient e-folds for sub-Planckian axion decay constants, moderate field ranges and inflationary scales, when non-perturbative effects are included. This puts them back into the favoured region of current observations and the weakly coupled, supergravity limit of string theory.  Such a scenario has distinctive signatures.  Whereas models with tiny modulations have large tensor modes, the benchmark models considered here predict tensor modes 2-4 orders of magnitude below\footnote{Tuning the parameters differently could easily increase the field ranges and tensor modes to within future bounds, approaching those models with tiny modulations \cite{AM1, Kobayashi:2010pz, Tatsuo, Kappl:2015esy,Choi:2015aem}.  However, as has already been emphasised, such models would call upon super-Planckian axion decay constants for bumpy natural inflation, and a scale of inflation close to GUT scale, which is difficult to realise consistently within perturbative string theory \cite{KPZ}.} the projected bounds of future observations like LiteBIRD, $r \lesssim 10^{-3}$.   Moreover, bumpy models also predict  large, negative values for the running of the scalar spectral index over the scales probed by the CMB ($\alpha_s \approx -10^{-2}, -10^{-3}$, respectively), and small running of the running ($\beta_s \approx -10^{-4}, -10^{-5}$, respectively).  This may be compared with a running that can oscillate between negative and positive in the case of small, frequent modulations \cite{Kobayashi:2010pz, Kappl:2015esy,Choi:2015aem}.  Additionally, previous models with small, frequent modulations have been used to motivate oscillatory features in the power spectrum  which can be searched for in the CMB \cite{AM1, AM3}, though recent analyses suggest such effects do not lie in the data \cite{PlanckMono}.  One would not expect to see this oscillatory behaviour in the bumpy models considered here, as the modulations have a much longer wavelength (the axion decay constant $f$ is larger) and the inflaton explores roughly only one period during the epoch of horizon crossing probed by the CMB.  To summarise, improvements in the measurements of $r$, $\alpha_s$ and $\beta_s$, could therefore distinguish or rule out the bumpy models. 

In this paper we considered simple potentials within effective field theory, and it would be important to understand whether the potentials and parameters emerging from string compactifications can fullfill the requisite tunings. Also, a more detailed study of the inflationary observables would be very interesting, for example consequences of the running spectral index on all the scales probed by the CMB, and implications of the bumps for non-Gaussianities.

\acknowledgements
It is a pleasure to thank Encieh Erfani for comments on the manuscript and Fernando Marchesano and Fernando Quevedo for discussions. 
The research of SLP is supported by a Marie Curie Intra European Fellowship within the 7th European Community Framework Programme. GT and IZ research is partially supported by STFC grants ST/N001435/1  and
ST/N001419/1 respectively. 

\section*{Appendix: Analytic study of a representative  case using the Hamilton-Jacobi approach}

We have seen in main the text, using numerical analysis,  that the Hubble friction due to the high velocity acquired rolling over the steep cliffs of the potential, is very effective in slowing down the inflaton when reaching the gentle plateaus. 
In this appendix, we discuss an explicit, analytical model that
 shares some of the interesting features of the scenarios we discussed. It
  allows us to solve the inflationary dynamics exactly, and have analytical control over  our findings.

We adopt the Hamilton-Jacobi approach \cite{Salopek:1990jq,Byrnes:2009qy}  reviewed in the main text, but for simplicity we set the Planck mass to one: $M_{ Pl}=1$.
We make the hypothesis that the Hubble parameter can be expressed as a function of the scalar field: this amounts to considering
only the effect of the `growing mode' for the dynamics \cite{Salopek:1990jq}. The homogeneous  evolution equations are
\bea
\dot \phi&=&-2 H'\label{pfoe}
\\
3 \,H^2&=&
V(\phi)+2\,  H'^2\label{sfoe}\,.
\eea
 Indeed, taking the time derivative of eq \eqref{pfoe}, and combining with the $\phi$-derivative
  of eq \eqref{sfoe}, one obtains the usual equations for a scalar in a FRW background.
 
We make the following, specific choice for the Hubble 
 parameter as a function of the scalar 
 field 
 \be
H\left( \phi \right)\,=\,m\,\phi+ \frac{m}{w}\,\sin{\left( w \,{\phi}  \right)} 
\,.
 \ee
 This is  a one parameter deformation -- with parameter $w$ -- of the Hubble parameter associated with 
 a model of
  $\phi^2$ inflation, which can be
 obtained when  $w\to0$: $H_{w\to0}\,=\,2\, m\, \phi$. 
 The effect of the contribution depending
on $w$ is to add bumps to the potential. 
 
 Indeed
 the corresponding potential during inflation, calculated by means of eq \eqref{sfoe}, is given by the expression
\bea
V(\phi)&=&3\, H^2-2\left( \partial_\phi H\right)^2\,,
\\
&=&3\,m^2\, \phi^2-2 \,m^2\nonumber\\
&&
+\frac{m^2}{w}\,\left[
6\,\phi\,\sin{\left( w \phi\right)}-4 w \cos{\left(w \phi \right)}
 \right]
\nonumber\\
&&+\frac{m^2}{w^2}\left[ 3 
\sin^2{\left( w \phi\right)} 
-2 w^2 \cos^2{\left( w \phi\right)} 
\right]\,.
\eea
 \begin{figure}[!htb!]
\begin{center}
\includegraphics[width=0.52\textwidth]{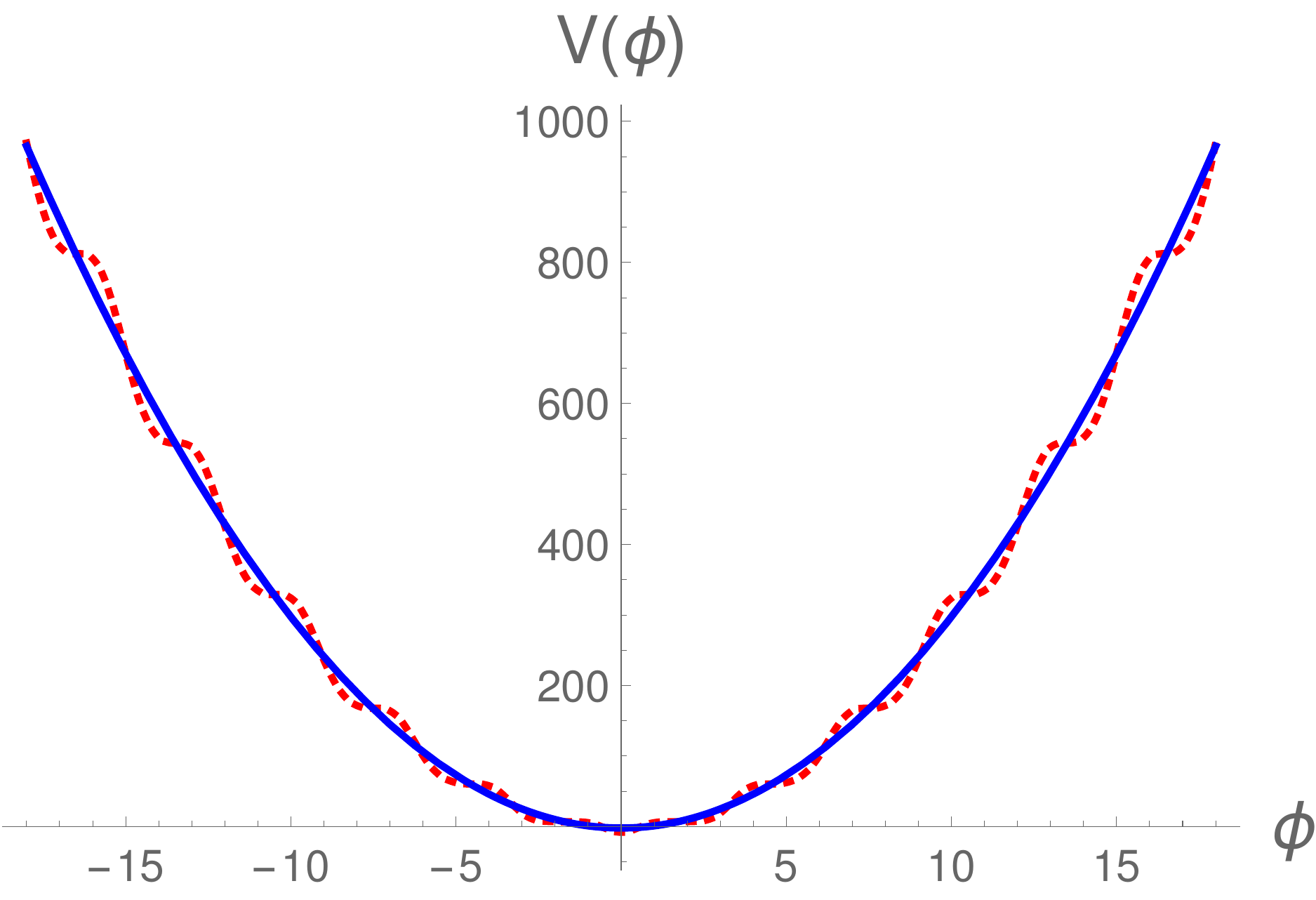} 
 \caption{{ A representation of the
potential for $m=2$,  $w=2.1$ (blue, solid) and the corresponding $\phi^2$ potential (red, dotted). }}\label{figp11}
\end{center}
\end{figure}  

Notice that the periodicity of the potential is broken by the mass parameter $m$. 
 See Fig \ref{figp11}
for a representative plot of the resulting potential, in comparison with a $\phi^2$ model.

\begin{figure}[!htb!]
\begin{center}
\includegraphics[width=0.42\textwidth]{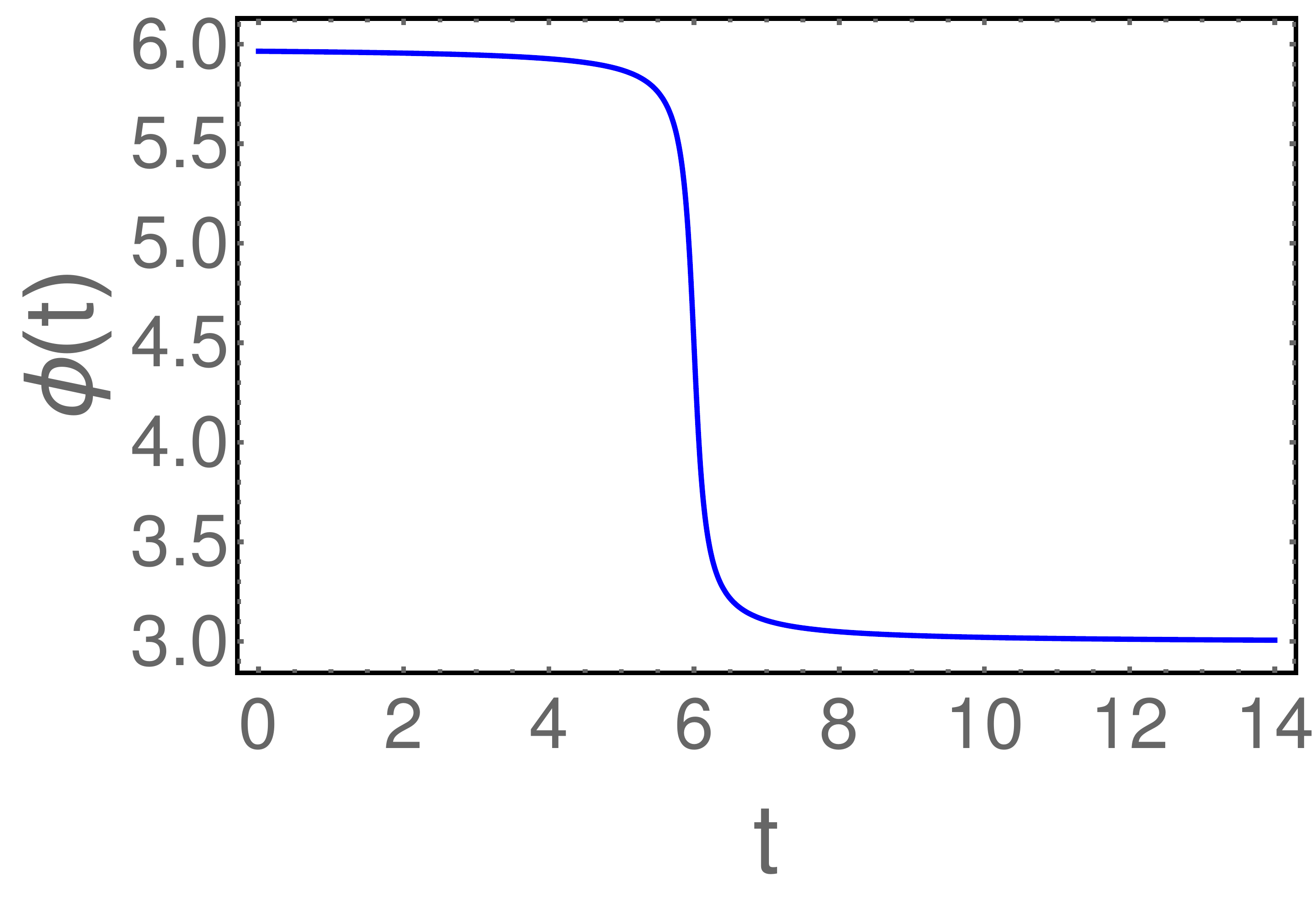} 
\includegraphics[width=0.42\textwidth]{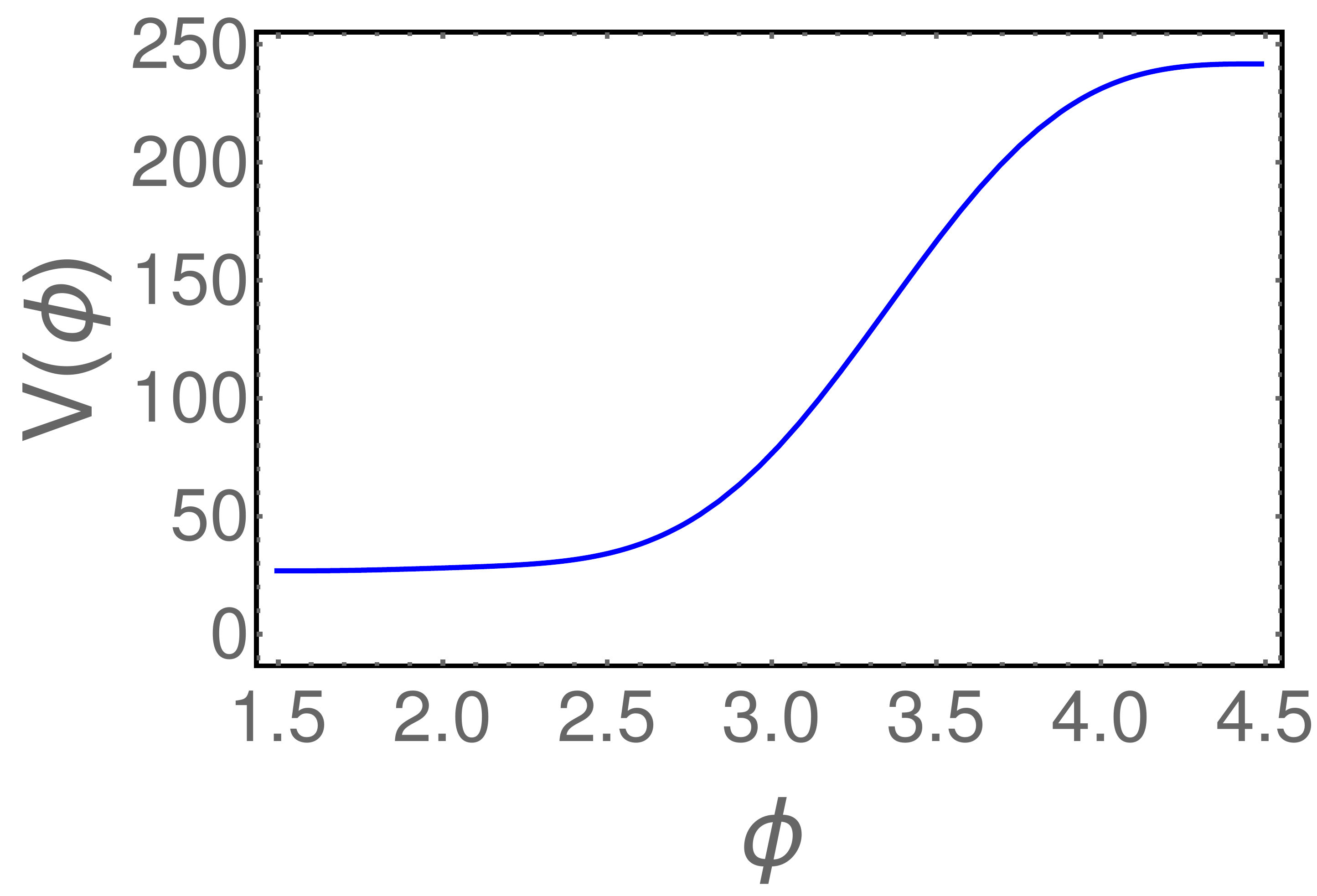} 
 \caption{{ {\bf Left} A representative example of the
scalar field solution $\phi(t)$ for $m=2$,  $w=2.1$, and $t_*=6$ and $n=2$.
{\bf Right} The
 single bump of the potential probed by the scalar trajectory of the left figure. }}\label{figp13}
\end{center}
\end{figure}

The dynamical evolution for the scalar field in this system reduces to
\be \label{eqts1}
\dot{\phi}+2m\left(1+\cos{w \phi}\right)\,=\,0 \,.
\ee
We can solve this equation  exactly. 
 First, one notices that the equation is `periodic', in the sense that it is  invariant under translations
\be
\phi\to\phi+\frac{2 \pi\, n}{w} \label{perr}
\ee
for an arbitrary integer $n$.  This implies that it is sufficient to study the solution in any of the
intervals
\be
\frac{(n-1)\pi}{w}\,\le\phi \le \frac{(n+1)\pi}{w} \label{phint}
\ee
for even values of $n$, and in the remaining regions the solution can be found using periodicity of eq \eqref{perr}. This suggests
to make the field redefinition from $\phi$ to $y$
\be
\phi(t)\,=\,\frac{n \pi}{w}+\frac{2}{w}\,\arctan{[y(t)]}
\ee
for even $n$, 
since one is then ensured that $\phi$ can probe  the entire  interval \eqref{phint}. Plugging the previous expression
into eq \eqref{eqts1}, one finds an equation  for $y$:
\be
\dot{y}+2 \,m \,w\,=\,0\,,
\ee
whose simple solution leads to the general solution for 
 $\phi$ 
 \bea\label{solph1}
\phi(t)&=&\frac{n \pi}{w}+\frac{2}{w}\arctan\left[ 2\,m\,w\left( 
t_*-t
\right)
\right]
\eea
where $t_*$ is an integration constant.  This solution corresponds to  a domain wall configuration for
$\phi$, whose dynamics in a finite time  is confined within only  one `bump' of the potential.  See Fig. \ref{figp13}
for a graphical representation of these findings. 
The resulting slope of the scalar field closely resembles the slopes we have met in the numerical analysis of the 
previous sections. Our interpretation is as follows: the scalar starts to roll very slowly in the plateau region of the potential, then it quickly
descends over the rapid cliff, and then Hubble friction slows down its motion, once it reaches the next plateau. This
analytical solution supports the numerical findings of our previous analysis.

\smallskip

This analytical
 model is so easy to handle that we can analytically compute the number of e-folds of inflation acquired during the inflationary trajectory. 
A simple calculation provides
the following expression
\bea
N_{ef} (w)&=&\frac{1}{2 w}\left(\phi_{in} \tan{\left[ \frac{w\,\phi_{in}}{2}\right]}
-\phi_{fin} \tan{\left[ \frac{w\,\phi_{fin}}{2}\right]}
\right)\,,
\eea
where $\phi_{in}$, $\phi_{fin}$ are  initial and final values of the scalar field, within the interval  \eqref{phint}, chosen
such that $\phi_{in}\ge \phi_{fin}$ (recall that the scalar starts up in the potential, and descends into the lower region).
  We can compare it to the number of e-folds obtained for the case with $w=0$, corresponding to $\phi^2$ inflationary model:
 \bea
N_{ef}(w=0)&=&\frac{1}{4}\left(\phi_{in}^2
-\phi_{fin}^2
\right)\,.
\eea
Let us choose for simplicity the simplest case where $n=0$ in the discussion above, and  $\phi_{fin}\,=\,0$. The ratio among the e-fold numbers
is
\be
{\text{ratio}}\,=\,\frac{N_{ef}(w)}{N_{ef}(w=0)}\,=\,\frac{\tan{\left[ {w\,\phi_{in}}/{2}\right]}}{\left(w \,\phi_{in}/2 \right)}\,.
\ee
This ratio is always greater than one, showing analytically that the deformation of the potential with a gentle plateau and a sharp cliff allows one to gain e-folds
of inflation. 

\bibliography{refs}

\bibliographystyle{utphys}

\end{document}